\documentclass[twocolumn, superscriptaddress, aps, prx]{revtex4-1}

\usepackage{graphicx}
\usepackage{amsmath}
\usepackage{amssymb}
\usepackage[hidelinks]{hyperref}

\graphicspath{{figures/}}
\setcitestyle{super}

\newcommand{\lession}[1]{$\blacksquare$~{\bf Lesson #1:}}
\renewcommand{\vec}[1]{\mathbf{#1}}
\DeclareRobustCommand*{\onlinecite}[1]{%
  \begingroup
    \romannumeral-`\x 
    \setcitestyle{numbers}%
    \cite{#1}%
  \endgroup
}

\begin{document}

\title{Next-generation non-local van der Waals density functional}

\author{D. Chakraborty}
\affiliation{Department of Physics, Wake Forest University,
Winston-Salem, NC 27109, USA.}
\affiliation{Center for Functional Materials, Wake Forest University,
Winston-Salem, NC 27109, USA.}

\author{K. Berland}
\affiliation{Faculty of Science and Technology, Norwegian University of
Life Sciences, Norway.}

\author{T. Thonhauser}
\email[E-mail: ]{thonhauser@wfu.edu}
\affiliation{Department of Physics, Wake Forest University,
Winston-Salem, NC 27109, USA.}
\affiliation{Center for Functional Materials, Wake Forest University,
Winston-Salem, NC 27109, USA.}

\date{\today}

\begin{abstract}
The fundamental ideas for a non-local density functional
theory---capable of reliably capturing van der Waals interactions---were
already conceived in the 1990's. In 2004, a seminal paper introduced the
first practical non-local exchange-correlation functional called vdW-DF,
which has become widely successful and laid the foundation for much
further research. However, since then, the functional form of vdW-DF has
remained unchanged. Several successful modifications paired the original
functional with different (local) exchange functionals to improve
performance and the successor vdW-DF2 also updated one internal
parameter. Bringing together different insights from almost two decades
of development and testing, we present the next-generation non-local
correlation functional called vdW-DF3, in which we change the functional
form while staying true to the original design philosophy. Although many
popular functionals show good performance around the binding separation
of van der Waals complexes, they often result in significant errors at
larger separations. With vdW-DF3, we address this problem by taking
advantage of a recently uncovered and largely unconstrained degree of
freedom within the vdW-DF framework that can be constrained through
empirical input, making our functional semi-empirical. For two different
parameterizations, we benchmark vdW-DF3 against a large set of
well-studied test cases and compare our results with the most popular
functionals, finding good performance in general for a wide array of
systems and a significant improvement in accuracy at larger separations.
Finally, we discuss the achievable performance within the current vdW-DF
framework, the flexibility in functional design offered by vdW-DF3, as
well as possible future directions for non-local van der Waals density
functional theory.
\end{abstract}

\maketitle

\section{Introduction}\label{sec:introduction}

Systems with van der Waals interactions are ubiquitous in nature and
they determine the structure of a vast and diverse array of materials
around us, reaching from cement to DNA. These materials are often of
scientific and technological importance, such as for gas storage and
sequestration,\cite{Tan_2016:trapping_gases,
Wang_2018:topologically_guided, Li_2017:capture_organic}
sensing,\cite{Jensen_2019:structure-driven_photoluminescence}
catalysis,\cite{Cure_2019:high_stability} organic
electronics,\cite{Diemer_2017:influence_isomer,
Gomez-Bombarelli_2016:design_efficient} and molecular crystals in
pharmaceutical,\cite{Reilly_2014:role_of_dispersion}
ferroelectric,\cite{Lee_2012:structure_energetics,
Shoji_2018:coputational_findings} and photovoltaic
applications.\cite{Kronik_2014:understanding_molecular_crystal,
Rangel_2016:structural_excited-state} It is therefore surprising that
capturing these interactions with standard materials modeling techniques
such as density functional theory (DFT) is still very challenging. Thus,
a major effort has been devoted to the inclusion of van der Waals forces
within DFT over the last two
decades.\cite{Grimme_2004:accurate_description,
Grimme_2007:density_functional, Grimme_2011:DFT_London_dispersion,
Tkatchenko_2009:accurate_molecular, Tkatchenko_2012:accurate_efficient,
Ambrosetti_2014:long_range_correlation, Ambrosetti_2016:wavelikeMBD,
Vydrov_2009:nonlocal_van, Vydrov_2010:dispersion_interactions,
Vydrov_2010:nonlocal_van, Grimme_2016:dispersion_corrected,
Szalewicz_2012:symmetry-adapted_perturbation,
Burns_2011:density-functional_approaches,
Klimes_2012:perspective_advances, Dion_2004:van_waals,
Thonhauser_2015:spin_signature, Langreth_2009:density_functional,
Berland_2015:van_waals} Within these developments, the non-local vdW-DF
family of functionals was a major breakthrough and stands out in that it
can be evaluated from knowledge of the density
alone.\cite{Dion_2004:van_waals, Langreth_2009:density_functional,
Berland_2015:van_waals,
Thonhauser_2015:spin_signature,Hyldgaard_2020:screening_nature} It
became popular because of its ability to provide accurate results for
binding energies and geometries of systems involving widely different
chemical compositions, ranging from typical van der Waals complexes to
adsorption on metallic surfaces.\cite{Langreth_2009:density_functional,
Berland_2014:van_waals, Berland_2015:van_waals} However, the emphasis in
vdW-DF's design has always been on accurately describing systems at
typical van der Waals separations, i.e.\ 3--4~\AA. As a result, errors
in interaction energies for larger---but yet still
relevant---separations often exceed the 100\%
mark.\cite{Lee_2011:evaluation_of_density_functional,
Hujo_2011:comparison_performance, gould_2018:are_dispersion_corrections}
With the more recent shift in research focus to truly extended systems
such as layered materials and surface adsorption, this issue becomes
highly pertinent.

The vdW-DF framework was published in 2004,\cite{Dion_2004:van_waals}
with the original functional form referred to here as vdW-DF1. Later
improvements\cite{Cooper_2010:van_waals, Klimes_2010:chemical_accuracy,
Wellendorff_2012:BEEF-vdW, Klimes_2011:van_waals,
Berland_2014:exchange_functional, Hamada_2014:van_waals} focused on
optimizing the local exchange with which vdW-DF is paired, while the
successor vdW-DF2 \cite{Murray_2009:investigation_exchange,
Lee_2010:higher-accuracy_van} also updated an internal parameter in the
non-local correlation part. All of these improvements provided essential
insight that informed the direction of further research and eventually
led to our current development. Overall, the vdW-DF family is remarkably
successful and widely used; considering the framework itself and all its
offsprings,\cite{Wellendorff_2012:BEEF-vdW, Cooper_2010:van_waals,
Thonhauser_2007:van_waals, Hamada_2014:van_waals,
Thonhauser_2015:spin_signature, Berland_2014:exchange_functional,
Lee_2010:higher-accuracy_van, Rydberg_2003:van_waals,
Klimes_2010:chemical_accuracy, Dion_2004:van_waals,
Klimes_2011:van_waals, Vydrov_2009:improving_accuracy} to date it has
received almost 12,000 citations.

However, all improvements of vdW-DF thus far have left its fundamental
framework unchanged since its inception. Here, we present an updated
framework for next-generation van der Waals density functionals. This
new framework is entirely built on the original
framework,\cite{Dion_2004:van_waals, Langreth_2009:density_functional}
which is rigorously derived from a many-body starting
point\cite{Rydberg_2001:thesis, Berland_2012:thesis,
Hyldgaard_2014:Interpretation_vdW, Berland_2015:van_waals,
Schroder_2017:vdw-df_family} and observes all the same constraints. In
our new development, we utilize a recently uncovered and largely
unconstrained degree of freedom in the underlying vdW-DF plasmon
dispersion model.\cite{Berland_2019:van_waals} This newly found
flexibility allows us to design a new functional form with two new
parameterizations that improve the performance at important mid-range
and larger separations \emph{without} sacrificing performance at binding
separations---overcoming this long-standing issue. We achieve this by
constraining this new degree of freedom in the plasmon dispersion model
through optimization to accurate quantum chemistry results for reference
systems. Our new non-local correlation functional form is a logical
extension and successor of the original
vdW-DF1\cite{Dion_2004:van_waals} and vdW-DF2-type
\cite{Lee_2010:higher-accuracy_van} functionals and hence we call it
vdW-DF3.

\section{Theory}\label{sec:Theory}

\subsection{Lessons Learned from Successive\\
Developments of vdW-DF}\label{sec:lessons}

The original vdW-DF1 of 2004 was of tremendous importance in
establishing the ability to describe van der Waals forces at the pure
DFT level. It introduced a non-local correlation energy functional of
the electron density $n(\vec{r})$ taking the form of a six-dimensional
integral
\begin{equation}
\label{equ:E_nlc}
E_{\rm c}^{\rm nl}[n] = \frac{1}{2}\int\! d^3 \vec{r} \int\! d^3 \vec{r'}\;
n(\vec{r}) \, \Phi(\vec{r}, \vec{r'}) \,n(\vec{r'}) \;,
\end{equation}
where the kernel $\Phi(\vec{r}, \vec{r'})$ connects different regions of
space and is derived from the adiabatic connection formula (ACF), see
Section~\ref{sec:background}. This non-local correlation energy
functional includes both short- and long-range contributions, but
vanishes seamlessly in the homogeneous electron-gas limit. In vdW-DF,
the non-local correlation part is therefore paired with that of the
local density approximation (LDA), $E_{\rm c}[n] = E_{\rm c}^{\rm
LDA}[n] + E_{\rm c}^{\rm nl}[n]$. The exchange part of vdW-DF, on the
other hand, is evaluated at the generalized-gradient level (GGA). The
GGA exchange can be expressed as a modulation of the LDA exchange as
\begin{equation}\label{equ:exchange}
E^{\rm GGA}_{\rm x}[n] = \int\! d^3 \vec{r} \; n(\vec{r})\,
\epsilon_{\rm x}^{\rm hom}\big(n(\vec{r})\big)\, F_{\rm x}(s)\;,
\end{equation}
where $\epsilon_{\rm x}^{\rm hom}$ is the exchange-per-particle in the
homogenous electron gas and the exchange enhancement factor $F_{\rm
x}(s)$ is a function of the reduced gradient $s(\vec{r}) \propto |\nabla
n(\vec{r})|/ n(\vec{r})^{4/3}$. In what follows, we briefly review
various vdW-DF developments and draw up a number of lessons learned from
them, which---in turn---influenced our functional design.

\begin{figure}[t!]
\includegraphics[width=\columnwidth]{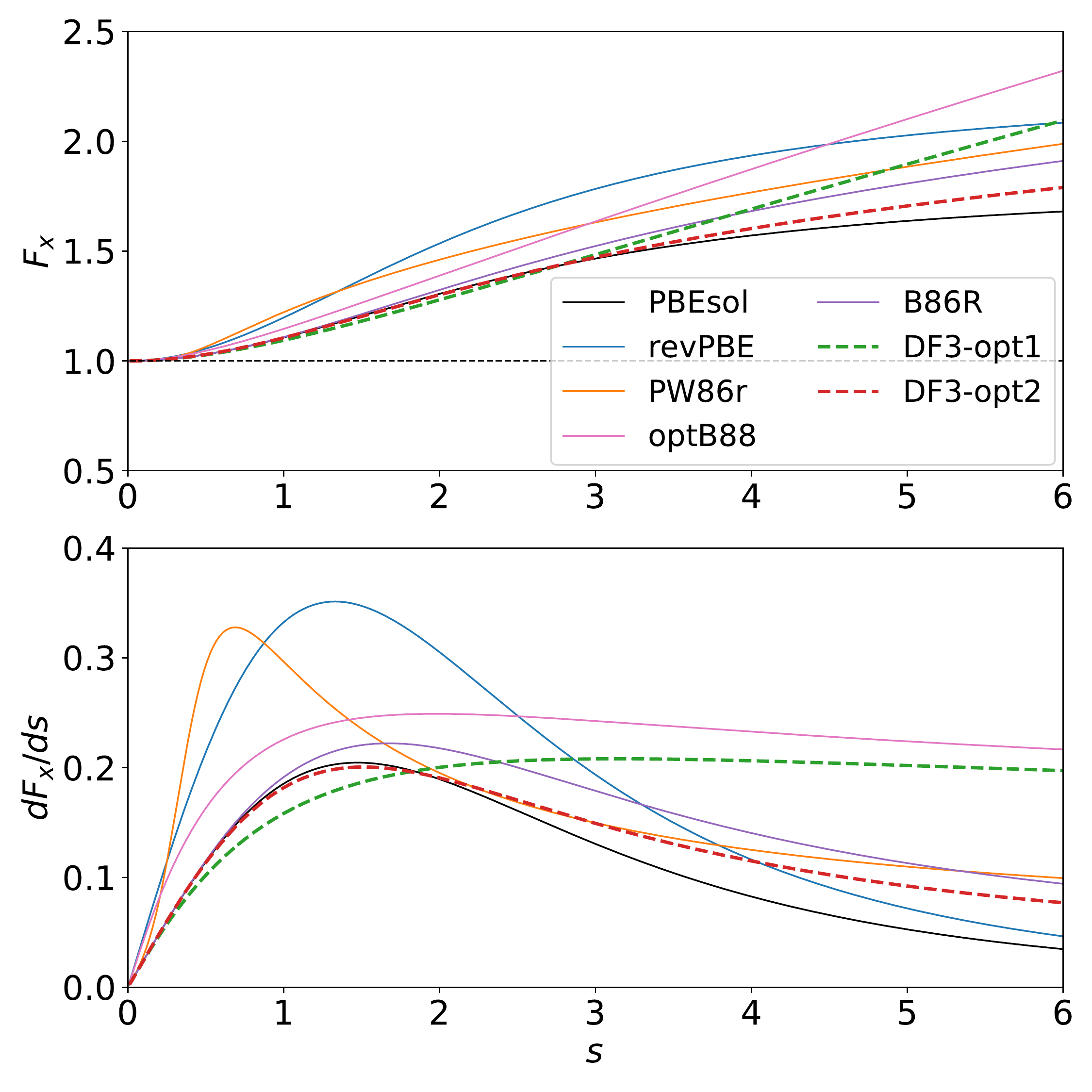}\\[-2ex]
\caption{\label{fig:Fx} {\bf(top)} Exchange enhancement factors $F_{\rm
x}(s)$ and {\bf(bottom)} their derivatives for selected functionals.}
\end{figure}

In vdW-DF1, revPBE exchange\cite{Perdew_1996:generalized_gradient_a,
Zhang_1998:comment_generalized} was chosen as the GGA exchange. This
choice was based on the fact that its rapidly increasing $F_{\rm x}(s)$
in the $s=0.5-2$ range, as shown in Fig.~\ref{fig:Fx}, ensures that
nonphysical binding effects from the exchange energy are kept at a
minimum.\cite{Dion_2004:van_waals, Murray_2009:investigation_exchange}
However, the choice of revPBE also leads to a consistent overestimation
of binding separations, occasionally causing incorrect bonding
predictions.\cite{Hamada_2010:comparative_vdW,
Yildrim_2013:trends_adsorption_characteristics, Liu2012:bz_metals,
Berland_2014:van_waals} After a number of studies had established both
the capabilities and shortcomings of
vdW-DF1,\cite{Chakarova-Kack_2006:application_vdW,
Cooper_2008:stacking_interactions, Kleis_2008:nature_strength,
Roman-Perez_2009:efficient_implementation,
Langreth_2009:density_functional, Toyoda_2009:first_principles_study}
the turn of the previous decade saw a string of important improvements.
First, Murray et al.\,\cite{Murray_2009:investigation_exchange}
demonstrated that a generally less rapidly but increasing $F_{\rm x}(s)$
for all values of $s$ (as well as in the asymptote) could also be used
to avoid spurious binding from the exchange energy. They did so by
reparameterizing the Perdew-Wang functional of
1986 (PW86r) \cite{Perdew_1986:accurate_simple} and showed that $F_{\rm
x}(s)\propto s^{2/5}$ for large values of $s$ is well suited to
reproduce the Hartree-Fock exchange interaction curves beyond binding
separations. This insight was used in the design of the successor
vdW-DF2 \cite{Lee_2010:higher-accuracy_van} which utilizes PW86r
exchange, but also changes an internal parameter from
$Z_{ab}^\text{DF1}= -0.8491$ to $Z_{ab}^\text{DF2}=-1.887$. This switch
effectively reduces the polarizability of a given density region, but
more so for highly inhomogeneous low-density regions than for high
density ones.\cite{Berland_2013:analysis_of_vdW} Through these changes,
vdW-DF2 obtains a significantly improved accuracy for molecular dimers;
however, for solids, layered systems, and some adsorption systems, the
development did not resolve the overestimation issues of vdW-DF1, which
in some cases even worsened.\cite{Klimes_2011:van_waals,
Berland_2015:van_waals, Bjorkman_2012:are_we_vdW_ready,
Tran_2019:nonlocal_vdW} This surprising worsening is possibly related to
the fact that the derivative of $F_{\rm x}(s)$ of PW86r is larger than
that of revPBE around $s=0.5$, see Fig.~\ref{fig:Fx}, as $d F_{\rm
x}(s)/d s$ is linked to the force exerted by the repulsive wall.

Around the same time, Cooper\cite{Cooper_2010:van_waals} demonstrated
that the systematic overestimation of binding separations could be
avoided by using a ``soft'' exchange functional, i.e.\ having an
exchange enhancement factor $F_{\rm x}(s)$ that increases slowly with
$s$ for small values of $s<1$. Similarly, the exchange functionals
optB86b\cite{Klimes_2011:van_waals} for vdW-DF1 correlation and
B86R\footnote{This functional is occasionally called
rev-vdW-DF2,\cite{Hamada_2014:van_waals} but we do not employ this
nomenclature, as it was only the exchange that was revised.} for vdW-DF2
correlation explicitly set the low-$s$ limit to that of the gradient
expansion $\mu_{\rm GEA}$ used in the soft
PBEsol,\cite{Perdew_2008:restoring_density-gradient} resulting in
significant improvements of lattice constants for solids. The various
enhancement factors and their performance for different systems have
been compared and analyzed in the context of vdW-DF in e.g.\
Refs.~\onlinecite{Murray_2009:investigation_exchange,
Berland_2015:van_waals, Schroder_2017:vdw-df_family}. To summarize, the
following was learned. \lession{1} \emph{The specific shape of $F_{\rm
x}(s)$ strongly impacts the bonding in vdW-DF and must be part of any
functional design. The small-$s$ limit should be soft, i.e.\ similar to
PBEsol \cite{Perdew_2008:restoring_density-gradient} in the small $s$
regime, to provide accurate solid lattice constants and $F_{\rm x}(s)$
should be asymptotically increasing (i.e.\ with a positive non-zero
derivative in the asymptotic $s$ limit) rather than going to a constant
to avoid spurious binding from the exchange energy.} This insight was
also used in the CX \cite{Berland_2014:exchange_functional} exchange
functional designed for vdW-DF1 correlation. While several GGA
functionals approach a constant value in the asymptote (such as PBE and
PBEsol) or for large $s$ do not exhibit monotonically and asymptotically
increasing $F_{\rm x}(s)$ (such as PW91\cite{PhysRevB.46.6671}), this
continued increase is essential within the vdW-DF framework to avoid
spurious binding stemming from the exchange part of the
exchange-correlation
functional.\cite{Murray_2009:investigation_exchange} We also note that
the analysis and discussion of $F_{\rm x}(s)$ as a function of $s$ has a
long history, predating the vdW-DF development.\cite{PhysRevA.47.4681,
doi:10.1063/1.475105, doi:10.1021/ct800522r}

Instead of updating the non-empirical criteria used in the design of
$F_{\rm x}(s)$, Klimes et al.\ \cite{Klimes_2010:chemical_accuracy}
fitted $F_{\rm x}(s)$ directly to the binding energies of the S22 data
set of molecular dimers keeping the vdW-DF1 correlation fixed for a set
of functional forms. These variants are therefore labeled as
semi-empirical or ``reference system optimized''. Their approach was
surprisingly effective in the sense that it not only improved binding
energies for molecular dimers, as would be expected, but also reduced
the overestimation of binding energies and improved performance for
several other classes of systems such as adsorption on coinage metals.
This is in particular the case for the optB88
\cite{Klimes_2010:chemical_accuracy} functional, which also arrived at a
quite soft (i.e.\ slowly increasing with $s$) small-$s$ from, but a
rapidly increasing high-$s$ form. This provides our next lesson.
\lession{2} \emph{Within vdW-DF, reference-system optimization to
specific benchmark sets has the potential to provide versatile
functionals.} A likely reason for this is the rigorously derived
non-local correlation model of vdW-DF, which is based on exact
constraints (see Section~\ref{sec:background}). The fitting to reference
data was also used successfully within the vdW-DF framework in the
construction of BEEF-vdW.\cite{Wellendorff_2012:BEEF-vdW} Fitting to
reference data is a common strategy in DFT functional development more
generally, done extensively for example in the popular Minnesota
functionals.\cite{C6SC00705H, doi:10.1021/acs.jctc.5b01082}

As of today, the optB88, optB86b, and CX exchange for vdW-DF1 and B86R
for vdW-DF2 are all actively used for broad classes of van der Waals
bonded materials and all have quite comparable overall performance, with
B86R possibly being slightly better for solids
\cite{Tran_2019:nonlocal_vdW} while optB88 is the only one providing
satisfactory results for rare gas dimers.\cite{Tran_2013:nonlocal_van}
\lession{3} \emph{It is not clear whether vdW-DF1 or vdW-DF2 correlation
is the best starting point for designing improved functionals, but in
any case a suitable exchange partner must be constructed once the
correlation functional is updated. In addition, the similar performance
of the best vdW-DF1 and vdW-DF2 variants indicate that tuning $Z_{ab}$
is not sufficient to greatly improve performance.} We also note that
both B88\cite{PhysRevA.38.3098} and B86b\cite{doi:10.1063/1.451353}
would be suitable starting points for reparameterizations of $F_{\rm
x}(s)$. This is less true for CX, as it was designed solely for the
vdW-DF1 correlation and is not as widely available in various codes,
though this is being remedied.\cite{Larsen_2017:libvdwxc_library}

Finally, in our recent work we found that tuning the momentum dependence
of the plasmon-pole model within vdW-DF provides an additional degree of
freedom that is fully consistent with the original constraint-based
design philosophy and that can be used to tailor various aspects of the
vdW-DF performance.\cite{Berland_2019:van_waals} In particular, we
learned two important points. \lession{4} \emph{The plasmon-pole model
is the key for improving the ability to simultaneously describe short-
and long-range contributions to van der Waals interactions and thus also
its ability to describe both small dimers and extended systems
accurately. And, the asymptotic behavior of any vdW-DF functional has
limited influence on the binding curves over physically relevant
distances.}

All these lessons laid the foundation for our design of vdW-DF3.

\subsection{Review of the Original vdW-DF Framework}\label{sec:background}

The kernel $\Phi(\vec{r}, \vec{r'})$ in Eq.~(\ref{equ:E_nlc}) can be
rigorously derived through a second-order expansion of the ACF. The
expansion is in terms of an effective plasmon propagator $S$, which
describes virtual charge-density fluctuations of the electron gas and
has poles for real frequencies at the effective plasmon frequency
$\omega_{\vec{q}}$, where $\vec{q}$ is the momentum of the
plasmon.\cite{Dion_2004:van_waals, Berland_2015:van_waals} Written
explicitly including the kernel $\Phi(\vec{r}, \vec{r'})$,
Eq.~(\ref{equ:E_nlc}) takes the form:
\begin{eqnarray}
\label{eq:vdW-DF}
E_{\rm c}^{\rm nl}[n] &=& \int_0^{\infty}\frac{du}{4\pi}\int\frac{d^3\vec{q}}
    {(2\pi)^3} \frac{d^3\vec{q}'}{(2\pi)^3}\times\nonumber\\[2ex]
&&  \big[1-(\hat{\vec{q}}\cdot\hat{\vec{q}}')^2\big]\; S_{\vec{q},\vec{q'}}
    ({\rm i}u) S_{\vec{q}',\vec{q}} ({\rm i}u)\;,
\end{eqnarray}
where $S$ is given by:
\begin{subequations}
\begin{eqnarray}
\label{eq:S1}
S_{\vec{q},\vec{q'}}({\rm i}u) &=&\frac{1}{2}\Big[\tilde{S}_{\vec{q},\vec{q'}}
   ({\rm i}u) + \tilde{S}_{-\vec{q'},-\vec{q}}({\rm i}u)\Big]\\\label{eq:S2}
\tilde{S}_{\vec{q},\vec{q'}}({\rm i}u) &=&\int\! d^3\vec{r}\;\frac{e^{-{\rm i}
   (\vec{q}-\vec{q}')\cdot\vec{r}}\;4\pi n(\vec{r})}{\big({\rm i}u +
   \omega_{\vec{q}}(\vec{r})\big)\big((-{\rm i}u + \omega_{\vec{q'}}(\vec{r})\big)}
\end{eqnarray}
\end{subequations}
Here, $u=-{\rm i}\,\omega$ is the imaginary frequency and $4\pi
n(\vec{r})$ is the square of the classical plasmon frequency. Note that
there are two symmetric two-point $S$ in Eq.~(\ref{eq:vdW-DF}), each of
which contains one density and one spatial integral (see
Eq.~(\ref{eq:S2})), leading to the two densities and spatial integrals
in Eq.~(\ref{equ:E_nlc}). This particular form of $S$ is chosen such
that it can fulfill four important physical constraints, i.e.\ time
invariance, charge conversation, the $f$-sum rule, and maintaining
self-correlation at large $q$.\cite{Dion_2004:van_waals,
Berland_2015:van_waals} These constrains are at the heart of vdW-DF and
make it a powerful and transferable tool for capturing van der Waals
interactions in vastly different systems.

As the main ingredient in Eq.~(\ref{eq:S2}), the dispersion model for
$\omega_\vec{q}$ comes into focus. The small-$q$ limit of
$\omega_\vec{q}$ has to be a constant (i.e.\ independent of $q$). On the
other hand, for the choice of $S$ in Eq.~(\ref{eq:S2}), the above
constraints are fulfilled if the plasmon dispersion has the large-$q$
limit of $\lim_{q\to\infty}\omega_\vec{q}(\vec{r}) = q^2/2$. For $q$
values in-between, the dispersion is not known. As such vdW-DF uses a
switching function $h$ that smoothly switches between the two known
limits. In particular, vdW-DF defines for the plasmon dispersion
\begin{equation}\label{equ:omega}
\omega_\vec{q}(\vec{r})=\frac{q^2}{2}\cdot\frac{1}{h\big(q/q_0(\vec{r})\big)}\;,
\end{equation}
where the switching function $h$ determines the relation between
density-density fluctuations and electromagnetic induction at different
length scales. To facilitate the numerical evaluation, the vdW-DF
framework uses only one length scale $\sim 1/q_0\big(n(\vec{r})\big)$ in
the switching function, which depends on the density and parameterizes
the local response of the electron gas. $q_0(\vec{r})$ is determined by
the requirement that the first-order expansion of the ACF in $S$
reproduces a generalized gradient approximation-type local
exchange-correlation (XC) functional. This XC functional is referred to
as the \emph{internal functional}, $\epsilon^{\rm int}_{\rm xc}$, and is
in general different from the total exchange-correlation functional. The
first-order expansion then yields for the internal
functional\cite{Dion_2004:van_waals}
\begin{align}
\epsilon^{\rm int}_{\rm xc}(\vec{r}) &= \pi \int \frac{d^3
     \vec{q}}{(2\pi)^3} \left[ \frac{1}{\omega_\vec{q}(\vec{r})}
     - \frac{2}{q^2} \right]  \nonumber \\
&= 2 \pi \int \frac{d^3 \vec{q}}{(2\pi)^3} \frac{1}{q^2}
     \left[ h\big(q/q_0(\vec{r})\big) -1 \right]\nonumber\\
&= -\frac{1}{\pi}\,q_0(\vec{r})\int_0^\infty d y \, [1 - h(y) ] \;.
\label{eq:eq1}
\end{align}
If we set
\begin{eqnarray}\label{eq:hycnstr}
\int_0^{\infty} dy\,[1- h(y)]=\frac{3}{4}\;,
\end{eqnarray}
then $q_0(\vec{r})$ takes a particularly simple form of a modulated
Fermi wave vector $k_{\rm F}^3(\vec{r})=3\pi^2n(\vec{r})$, i.e.\
$q_0(\vec{r})=-(4\pi/3)\,\epsilon^{\rm int}_{\rm
xc}(\vec{r})=\big(\epsilon^{\rm int}_{\rm xc}(\vec{r})/\epsilon_{\rm
x}^{\rm LDA}(\vec{r})\big)\,k_{\rm F}(\vec{r})$. For practical purposes,
the internal functional is approximated as LDA exchange correlation plus
simple quadratic exchange gradient corrections of the form $-
Z_{ab}^\text{DF1(2)} s^2/9$, which differ in vdW-DF1 and vdW-DF2 because
their $Z_{ab}$ are different. Both these functionals represent two
different directions for design philosophies and yield varying levels of
accuracy for different classes of
materials.\cite{Berland_2015:van_waals}

Obvious constraints on $h(y)$ are Eq.~(\ref{eq:hycnstr}) and that
$\lim_{y \to\infty}h(y)=1$ to fulfill the large-$q$ limit of
$\omega_\vec{q}(\vec{r})$. A third constraint, i.e.\ $h(0)=0$,
corresponds to charge conservation of the spherical XC hole model of the
internal functional.\cite{Hyldgaard_2014:Interpretation_vdW} The
original vdW-DF framework chooses a particular simple switching function
that fulfills all of those constraints trivially as $h_{\rm orig}(y)= 1
- \exp(-\gamma y^2)$, where $\gamma=4 \pi/9$. However, the three
constraints do still leave considerable freedom and more complicated
forms of $h$ are conceivable---yet, staying completely within the
original framework and thus inheriting its constraint-based
transferability.

As a final point, the dispersion model for $\omega_\vec{q}$ increases
with $q$, which effectively dampens the dispersion correction compared
to the asymptotic response. Thus, tuning the dispersion model is
somewhat reminiscent of optimizing damping functions in
dispersion-correction schemes that start from an asymptotic
formula.\cite{doi:10.1063/1.3382344} However, there are important
distinctions between damping in vdW-DF and dispersion-corrected methods
or semi-empirical non-local correlation functionals such as VV10: In
vdW-DF, the total non-local correlation has both short-range repulsion
and long-range attraction components and the wiggle-room in possible
dispersion models is highly constrained by the large-$q$ limit and the
integral in Eq.~(\ref{eq:hycnstr}).

\subsection{New Development}\label{sec:new_development}

We have recently demonstrated that the freedom in choosing the $h$
function can be exploited to significantly improve the notoriously bad
$C_6$ coefficients that derive from the vdW-DF
framework.\cite{Berland_2019:van_waals} From our work it became obvious
that this newly found freedom directly translates into a significantly
expanded design freedom (Lesson 4). Although our focus in
Ref.~\onlinecite{Berland_2019:van_waals} was on the asymptote, we
nonetheless gained some general insight into what aspects of $h$ lead to
what outcomes. In this regard, the fixing of the $C_6$ coefficients was
a simpler task, as they are proportional to $\lim_{y\to 0}h(y)/y^2$; the
problem of fixing the $C_6$ coefficients (asymptotic behavior) is thus
separable from improving the binding (short-range behavior) and a
relatively simple $h$ function is sufficient.

In our new development, we explore a larger space of $h$ functions in
order to improve the general accuracy for short, medium, and long
separations. This problem is vastly more complicated compared to the
$C_6$ coefficients as it does not separate and competing interests have
to be balanced. The accurate description of interactions beyond the
binding separation is important for e.g.\ surface-molecule vibrations
and related zero-point
energies\cite{Lee_2011:evaluation_of_density_functional,
Lee_2012:benchmarking_van} as well as inter-layer binding and surface
adsorption, for which it has been nicely demonstrated in
Ref.~\onlinecite{Berland_2013:analysis_of_vdW}---in particular for
non-planar molecules such as C$_{60}$.

Based on what we learned from our work on the $C_6$ coefficient,
combined with an extensive amount of trial and error, we identified a
new switching function which is both smooth and more flexible, in the
form of
\begin{equation}
\label{eq:h_genDF3}
h(y)= 1 - \frac{1}{1 + \gamma y^2 + (\gamma^2-\beta) y^4 + \alpha y^8}\;.
\end{equation}
$\alpha$, $\beta$, and $\gamma$ are adjustable parameters in this model,
albeit one of them is constrained by Eq.~(\ref{eq:hycnstr}); we describe
in Sec.~\ref{sec:optimization_scheme} how we determine the values of
those parameters with the help of an optimization scheme (Lesson 2).
This particular form of $h$ has a small-$y$ expansion of the form
\begin{equation}
\label{eq:h_genDF3-exp}
h(y)=\gamma y^2 - \beta y^4 + (2\beta\gamma-\gamma^3)y^6 + \dots
\end{equation}
or equivalently,
\begin{eqnarray}
\label{eq:genomega_DF3}
\lefteqn{\omega_\vec{q}(\vec{r})\sim y^2/h(y)=}\\\nonumber
&=& 1/\gamma + \beta y^2/\gamma^2 + (\beta^2/\gamma^3 - 2 \beta/\gamma + \gamma) y^4+\dots
\end{eqnarray}
This allows a clear interpretation of the parameters, as the $\gamma$
parameter sets the long-range van der Waals interactions, whereas the
$\beta$ parameter is the leading-order term causing damping of van der
Waals interactions at shorter ranges. Finally, the $\alpha y^8$ term
ensures that Eq.~(\ref{eq:hycnstr}) can be fulfilled without interfering
with the series expansions determining the long- and medium-range
behavior of the functional. The particular form of $h$ is in part
inspired by the so-called vdW-DF-09 from Vydrov and
Voorhis,\cite{Vydrov_2009:improving_accuracy} which does not fulfill
Eq.~(\ref{eq:hycnstr}), and was designed just prior to the release of
the more well-known VV09 and VV10.\cite{Vydrov_2009:nonlocal_van,
Vydrov_2010:nonlocal_van} Note that, while Eq.~(\ref{eq:genomega_DF3})
does not contain the exponential term of the original $h$ function, it
can be made into a form very similar to $h_{\rm orig} (y)$ in the more
relevant $0 < y < 2$ range.

The $h$ function in Eq.~(\ref{eq:h_genDF3}) provides an independent
parameter for the $y^2$ term in the series expansion of
$\omega_\vec{q}(\vec{r})$, Eq.~(\ref{eq:genomega_DF3}). This freedom can
be beneficial for fine-tuning the strength of the van der Waals
interactions in the mid-range, a few \AA\, away from the optimum binding
separations. However, when trying to minimize the error in interaction
energy of van der Waals complexes from binding distances to mid-range
and larger distances, we find the somewhat surprising result that the
optimal $\beta$ is close to 0, so that we actually approximate it with
$\beta=0$. This simplifies Eqs.~(\ref{eq:h_genDF3}) --
(\ref{eq:genomega_DF3}) and we thus define our $h$ function for vdW-DF3
as
\begin{equation}
\label{eq:h_DF3}
h_{\rm DF3}(y)= 1 - \frac{1}{1 + \gamma y^2 + \gamma^2 y^4 + \alpha y^8}\;,
\end{equation}
which leads to the small-$y$ expansions
\begin{eqnarray}
\label{eq:h_DF3-exp}
h_{\rm DF3}(y)    &=& \gamma y^2 -\gamma^3y^6 + \dots\\
y^2/h_{\rm DF3}(y) &=& 1/\gamma + \gamma y^4+\dots\label{eq:omega_DF3}
\end{eqnarray}
The quadratic term in $y^2/h_{\rm DF3}(y)$ is absent, which correspond
to the long-range limit of vdW-DF being well suited to describe the
entire long-to-mid range van der Waals interactions; at the same time
$\beta=0$ also allows a sharper damping of van der Waals interactions in
the mid-to-short range because of a larger $\alpha$ term, corresponding to a
faster increase of $h$ at $y\gtrsim1$. Note that in this context
``long-range'' in our design does not correspond to the asymptotic
limit, but rather corresponds to separations of about 5 -- 6~\AA\ beyond
the optimum separation.

\begin{figure}[t!]
\includegraphics[width=\columnwidth]{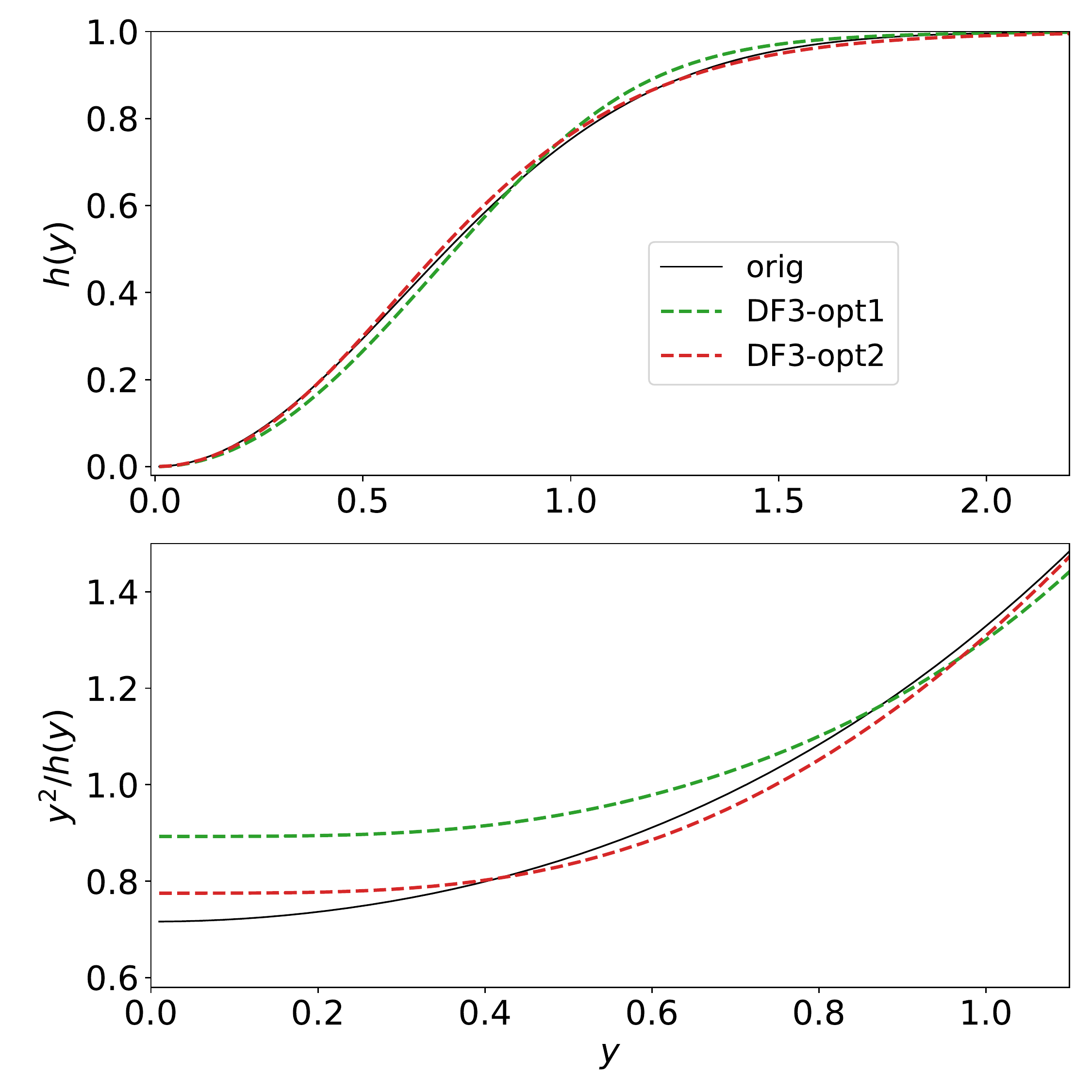}\\[-2ex]
\caption{\label{fig:h_function} {\bf(top)} Directly comparing the
various $h$ functions shows only minimal changes. {\bf(bottom)} Looking
at $y^2/h(y)$, which is proportional to the plasmon dispersion
$\omega_\vec{q}$, is much more revealing. The switching function $h_{\rm
orig}$ is taken from Ref.~\onlinecite{Dion_2004:van_waals}. Note the
different scales on the horizontal axes in both panels. The parameters
for $h_\text{DF3-opt1}$ and $h_\text{DF3-opt2}$ in Eq.~(\ref{eq:h_DF3})
are $(\alpha=0.94950, \gamma=1.12)$ and $(\alpha=0.28248, \gamma=1.29)$,
respectively.}
\end{figure}

Figure~\ref{fig:h_function} compares three different $h$ functions.
Although all these switching functions appear very similar when plotted
vs.\ $y$, a different picture emerges when plotting the physically
relevant quantity $y^2/h(y)$, which shows stark difference for $y<0.8$.
As both vdW-DF1 and vdW-DF2 correlation is in use in standard
functionals today and their performance is comparable (Lesson 3), for
our new functional form we want to explore possibilities for
improvements both within the vdW-DF1 \emph{and} vdW-DF2 design
philosophies and thus present two different parameterizations, which we
call $h_\text{DF3-opt1}$ and $h_\text{DF3-opt2}$. Both functions are
nearly constant for $y^2/h_{\rm DF3}(y)$ within $0<y<0.3$, which is
related to $\beta =0$. In contrast, for $h_{\rm orig}$ this function
behaves quadratic for small $y$. All plotted $h$ functions have
intercepts at different values $\lim_{y\to0}y^2/h(y)=1/\gamma$ because
they all have different values for $\gamma$. This intercept is directly
related to the asymptotic behavior of the functional and different
degrees of accuracy for the corresponding $C_6$ coefficients can thus be
expected.\cite{Berland_2019:van_waals}

Our new switching functions $h_\text{DF3-opt1}$ and $h_\text{DF3-opt2}$
constitute a significant change of the original vdW-DF framework. Any
such modifications require careful attention to rebalancing the exchange
part in Eq.~(\ref{equ:exchange}) (Lesson 1). As the exchange largely
determines the local screening effects that characterize the chemical
binding, we choose to rebalance it through a reparameterization of the
free parameters within the enhancement factors $F_{\rm x}$(s) of a
GGA-based exchange. Since $h_\text{DF3-opt1}$ and $h_\text{DF3-opt2}$
are noticeably different, they both need their own exchange
reparameterization. Based on the requirements of the $s$ dependence of
$F_{\rm x}$ (Lesson 1), we use
\begin{eqnarray}
\label{eq:Fx-b86r}
F_{\rm x}^\text{DF3-opt1}(s) &=& 1+\frac{\mu s^2}
    {1+\mu\,s\,\operatorname{arcsinh} (c\,s)/\kappa}\\
\label{eq:Fx-optb88}
F_{\rm x}^\text{DF3-opt2}(s) &=& 1+\frac{\mu s^2}
    {(1+\mu s^2/\kappa)^{4/5}}\;,
\end{eqnarray}
where $c = 2^{4/3} (3\pi^2)^{1/3}$. These exchange functionals are
inspired by optB88\cite{Klimes_2010:chemical_accuracy} and
B86R,\cite{Hamada_2014:van_waals} which have previously been paired
successfully with vdW-DF1 and vdW-DF2. To describe `weakly homogeneous'
systems, such as solids, layered structures and surfaces, we choose
$\mu=\mu_{\rm PBEsol}=10/81$ for both forms (Lesson 1). However, the
larger density-gradient region $ 1 < s < 4 $, which directly influences
the non-local binding regions, needs to be optimized with respect to our
new vdW-DF3 non-local functional, which we achieve through including
$\kappa$ in our optimization scheme in
Sec.~\ref{sec:optimization_scheme}. Figure~\ref{fig:Fx} shows the
differences in the various enhancement factors and their first
derivatives. In both cases, the enhancement factors and their
derivatives are reduced at larger gradients ($1 < s < 4$), indicating
that these semi-local exchange functionals become less repulsive at
higher density gradients compared to the original functional forms that
inspired them, i.e.\ optB88 and B86R. Finally, we note that both the
$F_{\rm x}(s)$ of DF3-opt2 has a shape that is quite similar to that of
B86R and $h(y)$ that is quite similar to that of the original vdW-DF,
indicating the suitability of B86R for the vdW-DF2 correlation.
DF3-opt1, on the other hand, has no such close similarity with previous
functionals.

\subsection{Optimization Scheme}\label{sec:optimization_scheme}

Our original theoretical development leaves three adjustable parameters,
i.e.\ $\beta$ and $\gamma$ from the proposed new switching function in
Eq.~(\ref{eq:h_genDF3})---we constrain $\alpha$ for every pair of
$\beta$ and $\gamma$ through Eq.~(\ref{eq:hycnstr})---and $\kappa$ from
the enhancement factor in Eqs.~(\ref{eq:Fx-b86r}) and
(\ref{eq:Fx-optb88}). Since there are two different enhancement factors
with possibly different values for $\kappa$, in principle we have to
perform two three-dimensional optimizations. We are using a
reference-system optimization (Lesson 2), where our parameters are
optimized with reference to high-level quantum chemistry (QC) results at
the CCSD(T) level of the S22$\times$5
dataset.\cite{Pavel_2010:comparative_study_s22x5} The quantity to be
minimized is the deviation of our calculated interaction energies from
the CCDS(T) reference for all 22 systems and all 5 separations. To avoid
making the optimization dominated by the large molecular dimers with
large binding energies, the target to be minimized should be a relative
rather than an absolute energy difference. In particular, we considered
the following two measures: mean absolute relative deviation (MARD) and
a differently weighted variant which we call weighted mean absolute
relative deviation (WMARD), defined as
\begin{eqnarray}\label{MARD}
{\rm MARD}   &=& \frac{1}{n}\sum_{\rm sep=1}^n {\rm MARD_{sep}}\hskip2em\\
\label{WMARD}
{\rm WMARD}  &=& \frac{1}{n}\sum_{\rm sep=1}^n {\rm WMARD_{sep}}
\end{eqnarray}
where
\begin{eqnarray}\label{MARD_sep}
{\rm MARD_{sep}} &=& \frac{1}{m}\sum_{\rm sys=1}^m
    |E^{\rm DFT}_{\rm sys, sep}  - E^{\rm QC}_{\rm sys, sep}|/
    E^{\rm QC}_{\rm sys, sep}\hskip2em\\
\label{WMARD_sep}
{\rm WMARD_{sep}}&=& \frac{1}{m}\sum_{\rm sys=1}^m
    |E^{\rm DFT}_{\rm sys, sep}  - E^{\rm QC}_{\rm sys, sep} |/
    E^{\rm QC}_{\rm sys, opt}
\end{eqnarray}
For the S22$\times$5 set used in our optimization we have $n=5$ and
$m=22$. Note that MARD puts the deviation in relation to the QC result
at that separation and thus treats all separations on the same basis.
However, when using MARD we found that the optimization equally weights
large percentage deviations at large separations, which, however, may on
the absolute scale only be in the sub-meV range---to the detriment of
performance around the binding separation. We thus weigh the deviation
by the interaction energy at the optimal separation, $E^{\rm QC}_{\rm
sys, opt}$ (where ``opt'' is the one separation out of the five for
which the interaction energy is largest), and optimize WMARD instead.

\begin{table*}[t!]
\caption{\label{tab:parameters} Optimum parameters for vdW-DF3-opt1 and
vdW-DF3-opt2. We set $\beta$ to zero and constrained $\alpha$ through
Eq.~(\ref{eq:hycnstr}), leaving only $\gamma$ and $\kappa$ as adjustable
parameters.}
{\begin{tabular*}{\textwidth}{@{}l@{\extracolsep{\fill}}ccccccccr@{}}\hline\hline
functional & $Z_{ab}$                     & $h$ function   & $\alpha$&  $\beta$ & $\gamma$ & exchange                    & form                            & $\kappa$ & $\mu$                      \\\hline
DF3-opt1   & $Z_{ab}^\text{DF1}= -0.8491$ & $h_\text{DF3}$ & 0.94950 &        0 & 1.12     & $F_{\rm x}^\text{DF3-opt1}$ & B88\cite{PhysRevA.38.3098}      &     1.10 & $\mu_{\rm PBEsol}$ = 10/81 \\
DF3-opt2   & $Z_{ab}^\text{DF2}=-1.887$   & $h_\text{DF3}$ & 0.28248 &        0 & 1.29     & $F_{\rm x}^\text{DF3-opt2}$ & B86b\cite{doi:10.1063/1.451353} &     0.58 & $\mu_{\rm PBEsol}$ = 10/81 \\\hline\hline
\end{tabular*}}
\end{table*}
%

The optimization is now performed on a grid for all three parameters,
where we use a coarse grid at first and later a finer grid around the
minimum. Note that each point in this three-dimensional space requires
$22\times 5 + 22\times 2 = 154$ (dimers + monomers) calculations, which
quickly becomes cost prohibitive. We thus decouple the exchange degree
of freedom from the $h$-function degrees of freedom and transform the
three dimensional optimization into a one-dimensional and
two-dimensional optimization. This can be achieved through performing
non-selfconsistent calculations and extracting the exchange energy as a
function of $\kappa$ (which is almost entirely independent of $\beta$
and $\gamma$) and the non-local correlation as a function of $\beta$ and
$\gamma$ (which also to a good approximation can be viewed as
independent of $\kappa$).\footnote{The exchange part depends on $\beta$
and $\gamma$ only through the changes in density in fully
self-consistent calculations. The same is true for the dependence of the
non-local correlation on $\kappa$.} The total energy of any point in the
three-dimensional space can then be reconstructed by adding the various
contributions on the fly to optimize WMARD. In the end, we verified all
our results with fully self-consistent calculations and our numbers
reported here in all tables and figures are the results of fully
self-consistent calculations. Although this approach constitutes a
tremendous reduction in computational effort, we still performed roughly
50,000 non-selfconsistent calculations.

As mentioned in the previous section, we found optimized $\beta$ values
that are a small positive number and zero for DF3-opt1 and DF3-opt2,
respectively, so we chose to set $\beta=0$ and thus reduce the amount of
parameters in our functionals down to two. Our optimized values for
$\alpha$, $\gamma$, and $\kappa$ are collected in
Table~\ref{tab:parameters}. It is conceivable that the global WMARD
minimum, in particular for DF3-opt2, might occur for negative $\beta$,
but this breaks formal constraints of the vdW-DF construction. Even
though $\beta$ came out to be zero, we chose to present our formalism
including $\beta$ as this additional parameter could be important in the
design of vdW-DF3 variants based on broader benchmark sets, or perhaps
in the construction of special-purpose functionals (in combination with
carefully selected exchange functionals) such as for the description of
molecular crystals\cite{Peng_2019:van_waals} or surface adsorption
processes of importance to
catalysis.\cite{doi:10.1021/acs.jpclett.6b01022}

\section{Computational Details}\label{sec:computational_details}

All our calculations were performed with the \textsc{quantum espresso}
(QE) package,\cite{Giannozzi_2017:advanced_capabilities} where we
modified the kernel generation routines to implement our new functionals
vdW-DF3-opt1 and vdW-DF3-opt2; these functionals are now available in
the latest official version of QE. We used PBE GBRV ultrasoft
pseudopotentials because of their excellent
transferability.\cite{Garrity_2014:pseudopotentials_high-throughput} The
wave-function and density cutoffs were set to $\sim$680~eV (50~Ryd) and
$\sim$8200~eV (600 Ryd), respectively. Self-consistent calculations were
performed with an energy convergence criterion of $\sim$1.36$\times
10^{-7}$ eV ($1 \times 10^{-8}$ Ryd) and, where applicable, a force
convergence criterion of $\sim$2.6$\times 10^{-5}$ eV/\AA\ ($1 \times
10^{-6}$ Ryd/Bohr) was used for structure relaxations. For all
calculations including metals/semiconductors a Gaussian smearing with a
spread of $\sim$100~meV (7.35~mRyd) was used. Benchmarking of our new
functionals has been done on the molecular dimer datasets S22$\times$5
and S66$\times$8, the G2--1 and G2RC sets, a set of solids, layered
structures, molecular crystals, and benzene adsorption on Cu/Ag/Au
surfaces. We compare the performance of our new functionals with other,
well-used dispersion-corrected exchange-correlation functionals such as
vdW-DF (vdW-DF1),\cite{Dion_2004:van_waals}
vdW-DF1-optB88,\cite{Klimes_2010:chemical_accuracy}
vdW-DF1-cx,\cite{Berland_2014:exchange_functional,
Berland_2014:van_waals} vdW-DF2,\cite{Lee_2010:higher-accuracy_van}
vdW-DF2-B86R,\cite{Hamada_2014:van_waals}
rVV10,\cite{Sabatini_2013:nonlocal_van} and
SCAN+rVV10,\cite{Peng_2016:versatile_vdW_SCAN-rVV10} and we use the
following corresponding short names in all tables and figures: DF1,
DF1-optB88, DF1-cx, DF2, DF2-B86R, VV, and SCAN+VV, respectively. For
the molecular dimers, we calculated all SCAN+VV values; for solids,
layered structures, and adsorption on coinage metals we took readily
available values from the literature, but for our molecular crystals we
found no published SCAN+VV data.

For calculations on the dimer sets, spurious interactions due to the
period boundary conditions in QE were minimized by padding dimers and
monomers with at least 15~\AA\ of vacuum. A list of 22 metals,
semiconductors, and ionic salts were also used as in
Ref.~\onlinecite{Klimes_2011:van_waals} except Li. A
$15\times15\times15$ $k$-point mesh was used for these periodic solids.
To calculate their lattice constants and cohesive energies, a
Birch-Murnaghan equation-of-state was used and the individual atom
energies were calculated in a box surrounded by at least 15~\AA\ of
vacuum. Results for cohesive energies and lattice constants are in
addition compared to PBE\cite{Perdew_1996:generalized_gradient} and
PBEsol.\cite{Perdew_2008:restoring_density-gradient} The reference data
on zero-point corrected experimental lattice constants and atomization
energies are taken from Ref.~\onlinecite{Klimes_2011:van_waals} and
references therein. Several layered structures were also considered.
Experimental structures were retrieved from the Inorganic Crystal
Structure Database (ICSD). Following the procedure in
Refs.~\onlinecite{Peng_2016:versatile_vdW_SCAN-rVV10,
Bjorkmann_2012:vdW_bonding_layered, Bjorkman_2014:testing_several},
these layered structures were relaxed along the inter-layer axis
($c$-axis) with $12\times12\times6$ $k$-points, keeping the $a$-lattice
constant at its experimental value. Inter-layer binding energies have
been calculated using single layers with fixed $a$-lattice constant and
with at least 12~\AA\ vacuum along the $c$-axis, using a $12\times
12\times 1$ $k$-mesh. The corresponding reference data is taken from RPA
calculations in Ref.~\onlinecite{Bjorkman_2014:testing_several} and
references therein. The molecular crystal dataset X23 was also studied.
Here, calculations were performed starting from structures provided in
Ref.~\onlinecite{Reilly_2013:understanding_role}, followed by an
optimization of all structural degrees of freedom. Finally, benzene
adsorption on the (111) surface of the coinage metals Cu, Ag, and Au
have also been used as a benchmark, using the reference data in
Refs.~\onlinecite{Berland_2009:rings_sliding,
Liu_2013:structure_energetics, Bilic_2006:adsorption_benzene,
Maass_2018:binding_energies, Peng_2016:versatile_vdW_SCAN-rVV10}. Six
layers were used to form the metallic
slab,\cite{Berland_2009:rings_sliding} keeping the three bottom layers
fixed and using a 9~\AA\ vacuum. Calculations were performed with a
$4\times4\times1$ $k$-mesh.

\section{Results}\label{sec:results}

To investigate the performance of vdW-DF3-opt1 and vdW-DF3-opt2, we
benchmark those functionals on an extensive list of systems reaching
from molecular dimers to periodic systems including solids, layered
systems, molecular crystals, and surface adsorption on coinage metals.
We compare our results with the most popular functionals, finding good
performance in general for a wide array of systems and a significant
improvement in accuracy at larger separations.

\begin{figure*}
\includegraphics[width=\textwidth]{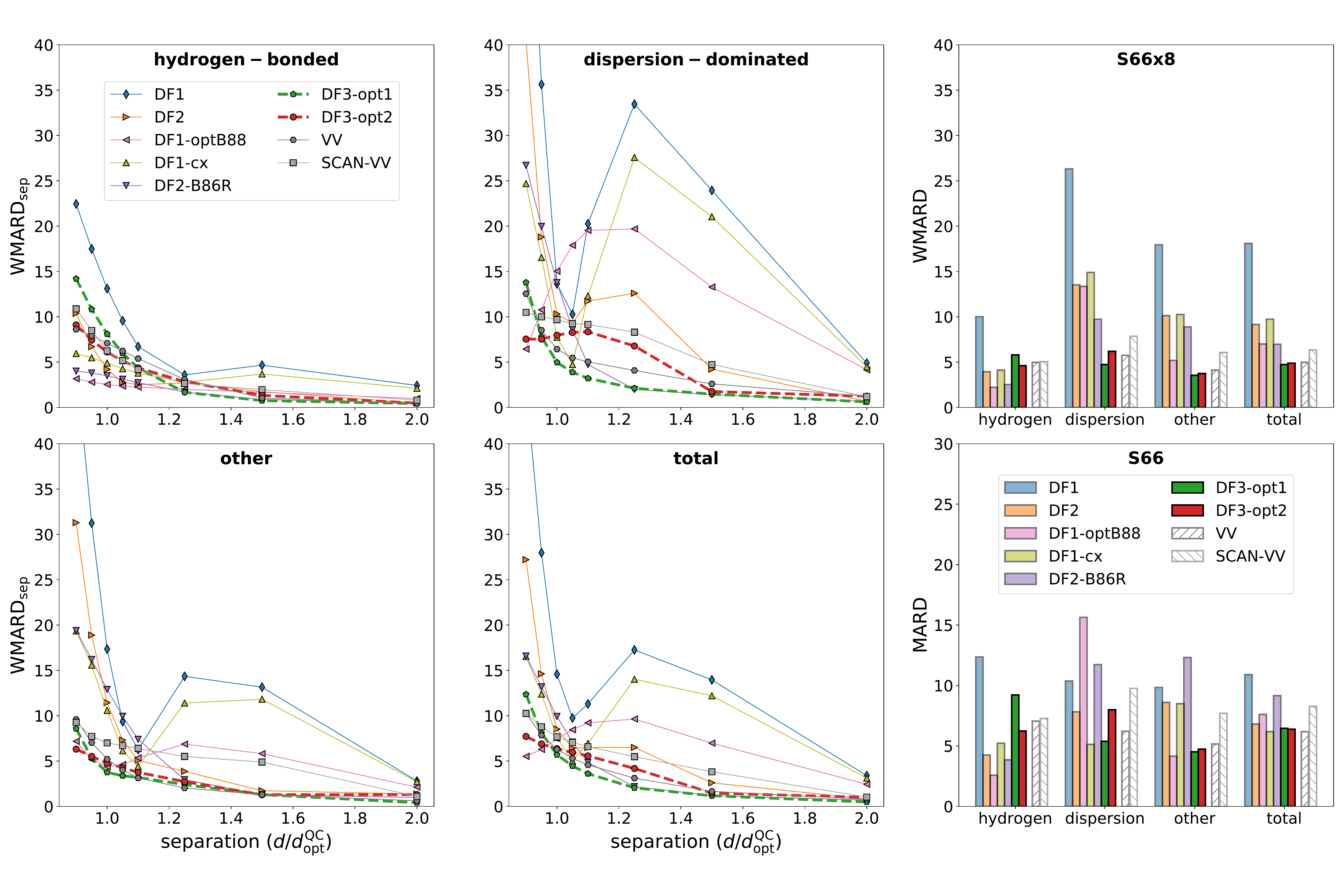}\\[-2ex]
\caption{\label{S66x8}Statistical analysis for the S66$\times$8 and S66
datasets in reference to QC data at the CCSD(T) level. The four left
figures show WMARD$_{\rm sep}$ from Eq.~(\ref{WMARD_sep}), summed over
different S66$\times$8 subgroups. Separation is given in units of the
optimal QC separation $d^{\rm QC}_{\rm opt}$. The top right panel shows
WMARD from Eq.~(\ref{WMARD}). The bottom right plot shows MARD for the
optimal binding energy of the S66 dataset. We compare our results also
to VV and SCAN+VV, but we separate them as they are fundamentally
different approaches and should not be understood as improvements within
the vdW-DF family.}
\end{figure*}

\begin{figure*}
\includegraphics[width=\textwidth]{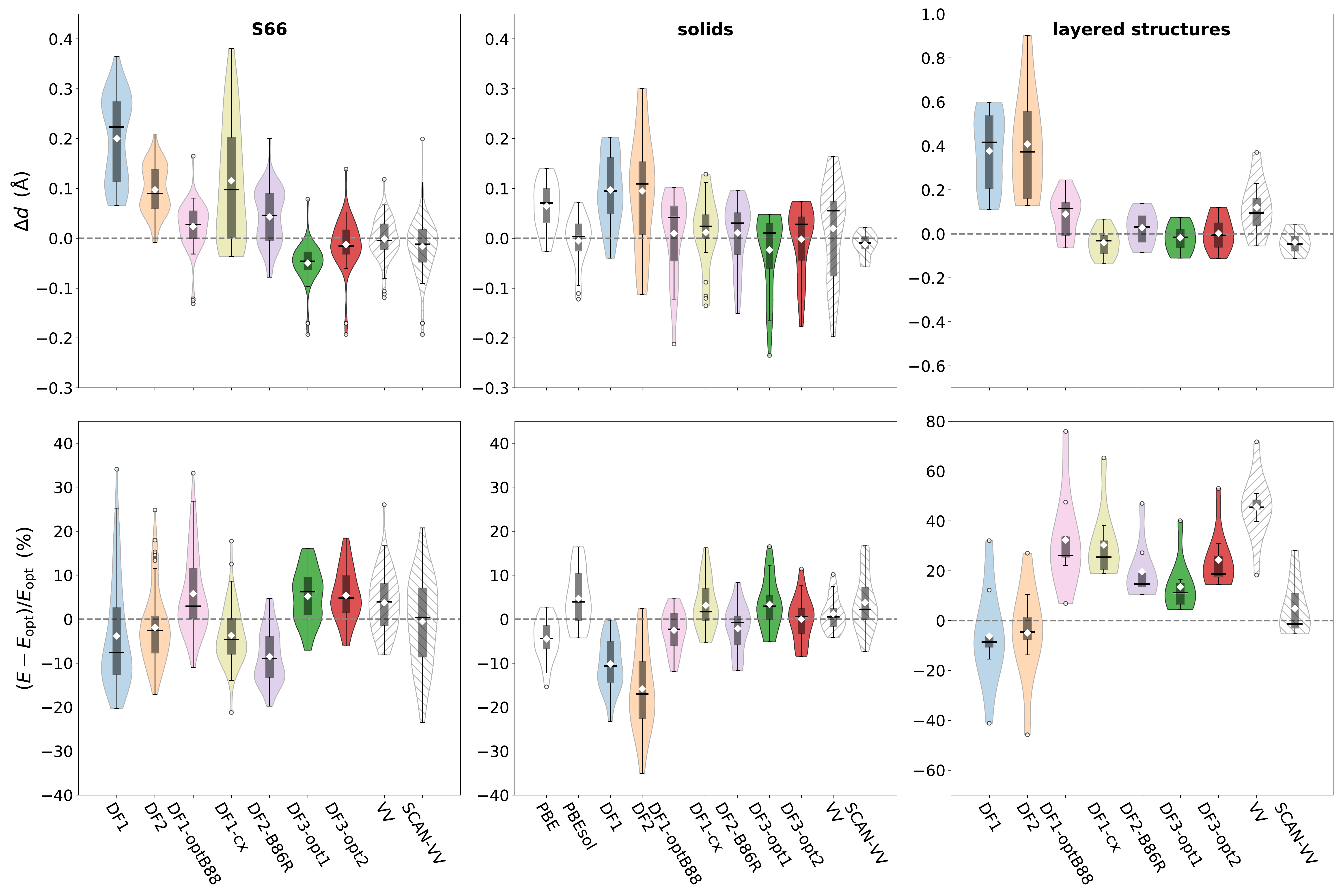}\\[-2ex]
\caption{\label{solid-violin-box} Violin plots overlaid on box plots of
the deviations from reference data for the different functionals. Violin
plots represent the data distribution and are based on a Gaussian kernel
density estimation using the Scott’s rule,\cite{Scott_1979:on_optimal}
as implemented in \textsc{matplotlib}. In the box plot, the boxes hold
50\% of the data, with equal number of data points above and below the
median deviation (full black line). Whiskers indicate the range of data
falling within 1.5~$\times$~box-length beyond the upper and lower limits
of the box. Outliers beyond this range are indicated with circular
makers. Diamonds mark the mean deviation. {\bf(left)} Set of 66
molecular dimers: reference data taken from
CCSD(T);\cite{Rezac_2011:s66_well-balanced} results in terms of $d$
(dimer separation) and $E_{\rm opt}$ (binding energy). {\bf(middle)} Set
of 22 solids: reference data taken from zero-point corrected
experiments;\cite{Csonka_2009:assessing_performance} $d$ refers to
lattice constant and $E_{\rm opt}$ to atomization energy. {\bf(right)}
Set of 9 layered structures: reference data taken from RPA
calculations;\cite{ Bjorkmann_2012:vdW_bonding_layered,
Bjorkman_2014:testing_several} $d$ refers to layer separation and
$E_{\rm opt}$ to layer binding energy. SCAN+VV data for solids and
layered structures taken from
Ref.~\onlinecite{Peng_2016:versatile_vdW_SCAN-rVV10}.}
\end{figure*}

\subsection{Molecular Systems}\label{subsec:molec-dimer}

The two adjustable parameters of our functionals (see
Table~\ref{tab:parameters}) have been fitted to minimize the WMARD of
the S22$\times$5 dataset,\cite{Pavel_2010:comparative_study_s22x5} as
described in Sec.~\ref{sec:optimization_scheme}. A comparison for this
dataset is thus biased by construction and we will not go into extensive
details here. Appendix~\ref{sec:statistics} holds a statistical summary
and detailed results for each dimer are provided in the Supporting
Information. Overall, both our new functionals have a WMARD of less than
4\% and perform best in our comparison group. The performance is
particularly good for dispersion-dominated complexes. Even though we
optimized WMARD, MARD also shows significant improvements.

The more diverse and larger S66$\times$8 set of molecular dimers is our
first proper benchmark.\cite{Rezac_2011:s66_well-balanced} Similar to
S22$\times$5, this set is comprised of 23 hydrogen bonded complexes, 23
dispersion-dominated complexes, and 20 complexes with various other
kinds of interactions. Interaction energies at the CCSD(T) level are
reported for eight different separations---two at separations below the
optimal binding distance, one at the optimal binding distance, and five
separations that are larger, up to twice the optimal binding separation.
The WMARD defined in Eq.~(\ref{WMARD}) for the S66$\times$8 set is given
in the upper right panel of Fig.~\ref{S66x8}; a summary of statistical
information can be found in Appendix~\ref{sec:statistics} and detailed
results for each dimer are provided in the Supporting Information. As
S66$\times$8 is quite similar to the S22$\times$5 set, our two
functionals also here perform best with a WMARD of 4.7\% and
4.9\%, although it has gone up by approximately one percentile. For
dispersion dominated and mixed complexes DF3-opt1 performs better than
all other tested with a WMARD of 4.7\% and 3.5\%, respectively. DF3-opt2
has slightly higher WMARD for dispersion-dominated system (6.2\%), but
is also very accurate (3.7\%) for mixed complexes.

A more detailed picture of the performance for the S66$\times$8 emerges
in Fig.~\ref{S66x8}, which provides WMARD$_{\rm sep}$ from
Eq.~(\ref{WMARD_sep}), summed over all three subgroups as well as for
all 66 complexes. The plots reveal that both DF3-opt1 and DF3-opt2
accurately describe interaction energies at equilibrium separation and
beyond for each interaction type. In particular, we consider the
``dispersion-dominated'' panel amongst the most pertinent results of our
study. It shows that DF3-opt1, and to a somewhat lesser extent DF3-opt2,
agrees well with the quantum-chemical reference data for
dispersion-bound systems beyond equilibrium separations---whereas
several popular functionals give quite large errors in this regime---and
thus confirms that we have achieved our goal of overcoming this
longstanding problem. For hydrogen-bonded systems DF3-opt1 turns out to
be less accurate than DF1-cx, DF2-B86R, and DF1-optB88 around the
binding separation, but still has good performance similar to SCAN+VV
and VV and shows a significant improvement over vdW-DF1 and
PBE+D3.\cite{doi:10.1002/cphc.201100826} DF3-opt2 shows an accuracy
quite similar to DF3-opt1, but with somewhat better performance for
hydrogen-bonded systems and short separations, at the cost of lower
accuracy for dispersion-dominated systems. The reason that DF3-opt1 (and
to a lesser extent DF3-opt2) is less accurate for hydrogen-bonded
systems around the binding separation may be related to the smaller
$dF_{\rm x}(s)/ds$ at around $s\approx 0.5-2$ compared to e.g.\ B86R or
optB88.\cite{Berland_2014:exchange_functional} Our analysis of $s$
values shows that this range is the most relevant for binding in
hydrogen-bonded molecular dimers. Furthermore, a comparison of various
forms of $F_{\rm x}(s)$ and $dF_{\rm x}(s)/ds$ and their influence on
performance for different systems and types of interactions has been
done in e.g.\ Refs.~\onlinecite{Murray_2009:investigation_exchange,
Berland_2015:van_waals, Schroder_2017:vdw-df_family}. In addition, in
Section~\ref{sec:balance} we provide further discussion on the inherent
trade-offs in vdW-DF design and give arguments that increased values of
$dF_{\rm x}(s)/ds$ in this range can improve performance for
hydrogen-bonded systems.

In addition, we also provide data for the S66 data
set,\cite{Rezac_2011:s66_well-balanced} which contains the same
molecular dimers as the S66$\times$8 set but uses the optimal binding
separation rather than looking at eight explicit separations. Thus, in
our comparison, we also fully optimize the binding separation with the
various functionals. The MARD of the resulting optimized binding
energies is given in the bottom right panel of Fig.~\ref{S66x8} and
statistical data for the deviations in optimal binding separation and
binding energy are analyzed in the left column of
Fig.~\ref{solid-violin-box} in the form of violin plots and box plots;
additional data is available in the Supporting Information. Again, we
find that DF3-opt1 and DF3-opt2 perform very well. In particular, the
violin plots reveal that our new functionals provide rather compact
results with less spread in comparison to other functionals.

As a simple check that our reparameterization of the non-local
correlation and semilocal exchange does not cause vdW-DF3 to fail in
basic chemistry areas, we test our two functionals for the
\mbox{G2--1}\cite{doi:10.1063/1.460205, doi:10.1063/1.473182} set of
molecular atomization energies as well as the
G2RC\cite{doi:10.1063/1.473182} set for reaction energies. Results are
depicted in Fig.~\ref{fig:G2-1}; additional information is available in
Appendix~\ref{sec:statistics} and the Supporting Information. Most
functionals perform similar for both tests---typically comparable to PBE
as expected\cite{doi:10.1063/1.4986522} and better than PBEsol---with
small deviations in spread and mean/median.

\begin{figure}
\includegraphics[width=0.95\columnwidth]{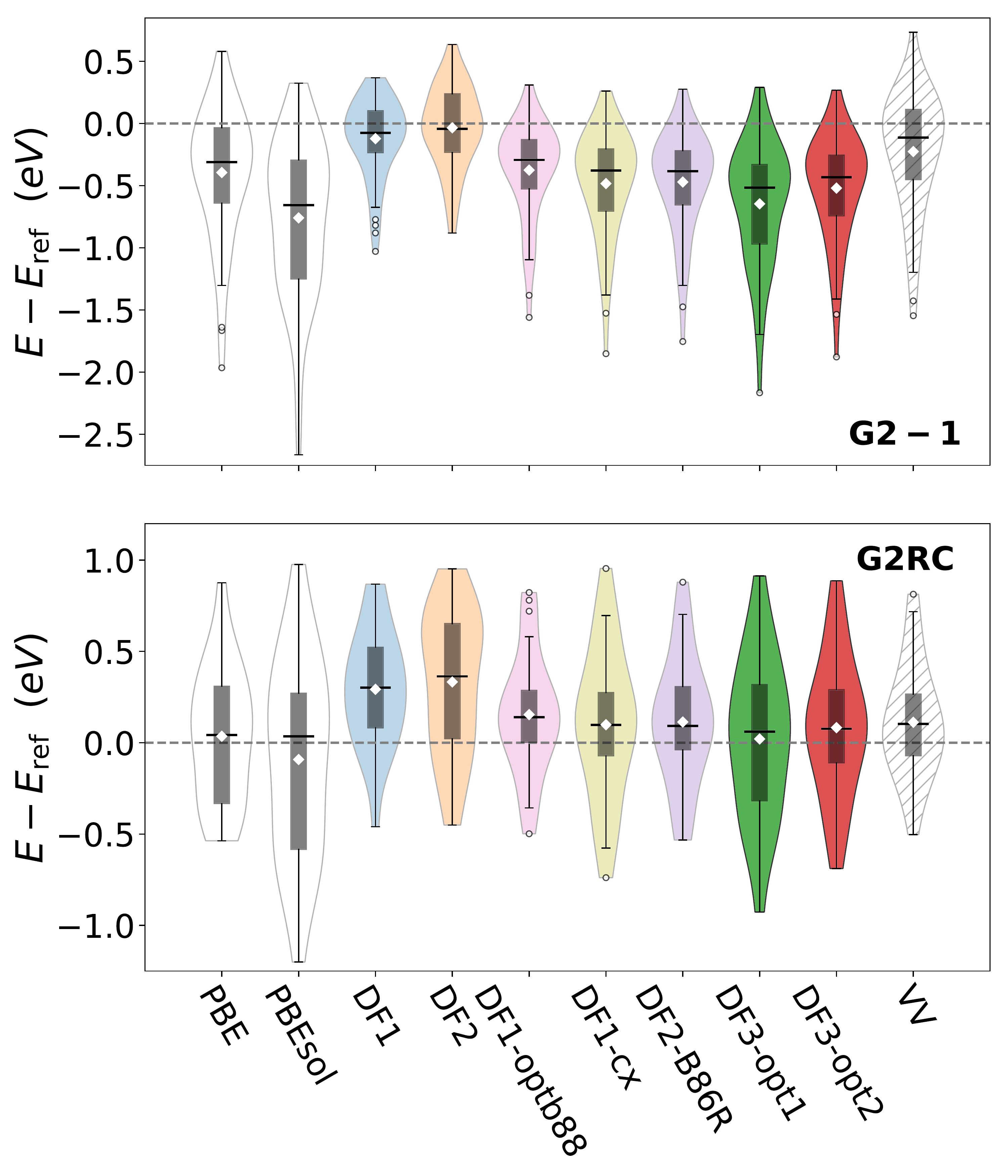}\\[-3ex]
\caption{\label{fig:G2-1}Violin plots of deviations for the atomization
energies of the G2--1 set and reaction energies of the G2RC set.
Reference data taken from Refs.~\onlinecite{Grimme_website,
doi:10.1063/1.1926272}. See caption of Fig.~\ref{solid-violin-box} for
further details.}
\end{figure}

\subsection{Solids}\label{sec:solid}

Within DF3-opt1 and DF3-opt2 the non-local correlation is purposefully
combined with an exchange energy that has a smaller, PBEsol-like
enhancement factor for small $s$, i.e.\ $F_{\rm
x}^\text{DF3-opt}(s)=1+\mu_{\rm PBEsol}s^2+\dots$, which significantly
improves lattice constants of solids. In Fig.~\ref{solid-violin-box} we
collect statistical information in the form of violin plots combined
with box plots for a set of 22 standard solids \footnote{We chose this
set as the intersection of the sets given in
Refs.~\onlinecite{Klimes_2010:chemical_accuracy, Klimes_2011:van_waals,
Peng_2016:versatile_vdW_SCAN-rVV10}} and provide deviations for lattice
constants and atomization energies. As reference we use results from
zero-point corrected
experiments.\cite{Csonka_2009:assessing_performance} Further numerical
data is provided in Appendix~\ref{sec:statistics}. Clearly, PBEsol and
SCAN+VV provide an accurate description of lattice constants. However,
DF3-opt1 and DF3-opt2, together with other recent functionals also show
good performance. In terms of atomization energies, we find several
functionals that perform well and even better than PBE, including our
new functionals. In particular, DF3-opt2 has a mean and median deviation
of essentially zero. Within the vdW-DF family of functionals, DF3-opt1
and DF3-opt2 retain this significant advancement in vdW-DF design, as
the original functionals DF1 and DF2 both overestimate lattice constants
for solids.

\begin{figure}
\includegraphics[width=0.9\columnwidth]{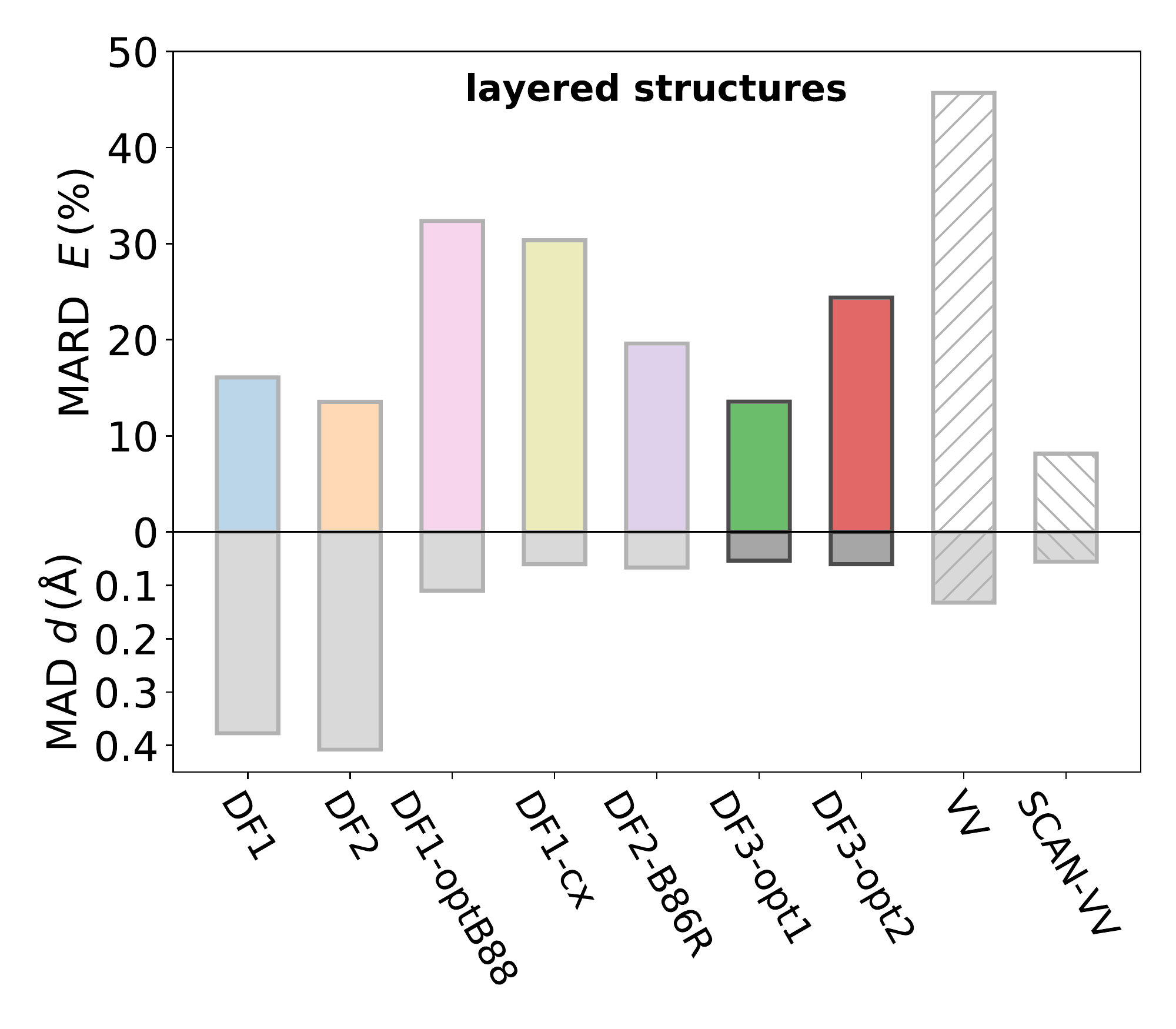}\\[-3ex]
\caption{\label{layered-str-MAD-plot}MARD of layer binding energy and
MAD of layer separation for a set of layered structures. SCAN+VV data
taken from Ref.~\onlinecite{Peng_2016:versatile_vdW_SCAN-rVV10}.}
\end{figure}

\subsection{Layered Structures}\label{sec:layered}

We also benchmark our functionals for a set of 9 layered structures
against RPA reference
calculations\cite{Bjorkmann_2012:vdW_bonding_layered,
Bjorkman_2014:testing_several} and results are given in the right column
of Fig.~\ref{solid-violin-box}. Futher details are provided in
Appendix~\ref{sec:statistics}, also see
Ref.~\onlinecite{Tran_2019:nonlocal_vdW}. While the original DF1 and DF2
significantly overestimate the layer separation, much improvement can be
seen for all other vdW-DF functionals. In particular, DF3-opt2 has a
mean deviation of zero and a compact spread, closely followed by
DF3-opt1. Improvements for the layer binding energy are mostly observed
in smaller spreads for newer vdW-DF functionals. While SCAN+VV is
remarkably accurate for these systems, DF3-opt1 performs best out of all
vdW-DF functionals. The progress made by our two functionals within the
vdW-DF family can better be seen in Fig.~\ref{layered-str-MAD-plot},
where we show the MARD of layer binding energy and MAD of layer
separation. The original DF1 and DF2 functionals had a reasonable MARD
for the energy, but their MAD in layer separation rendered them
inapplicable for layered structures. Further developments like
DF1-optB88, DF1-cx, and DF2-B86R corrected that behavior, but to the
detriment of MARD in energy. DF3-opt1 now noticeably reduces the MARD in
energy again (and also the spread, see Fig.~\ref{solid-violin-box})
while having the lowest MAD in layer separation of any tested
functional.

\begin{figure}
\includegraphics[width=0.9\columnwidth]{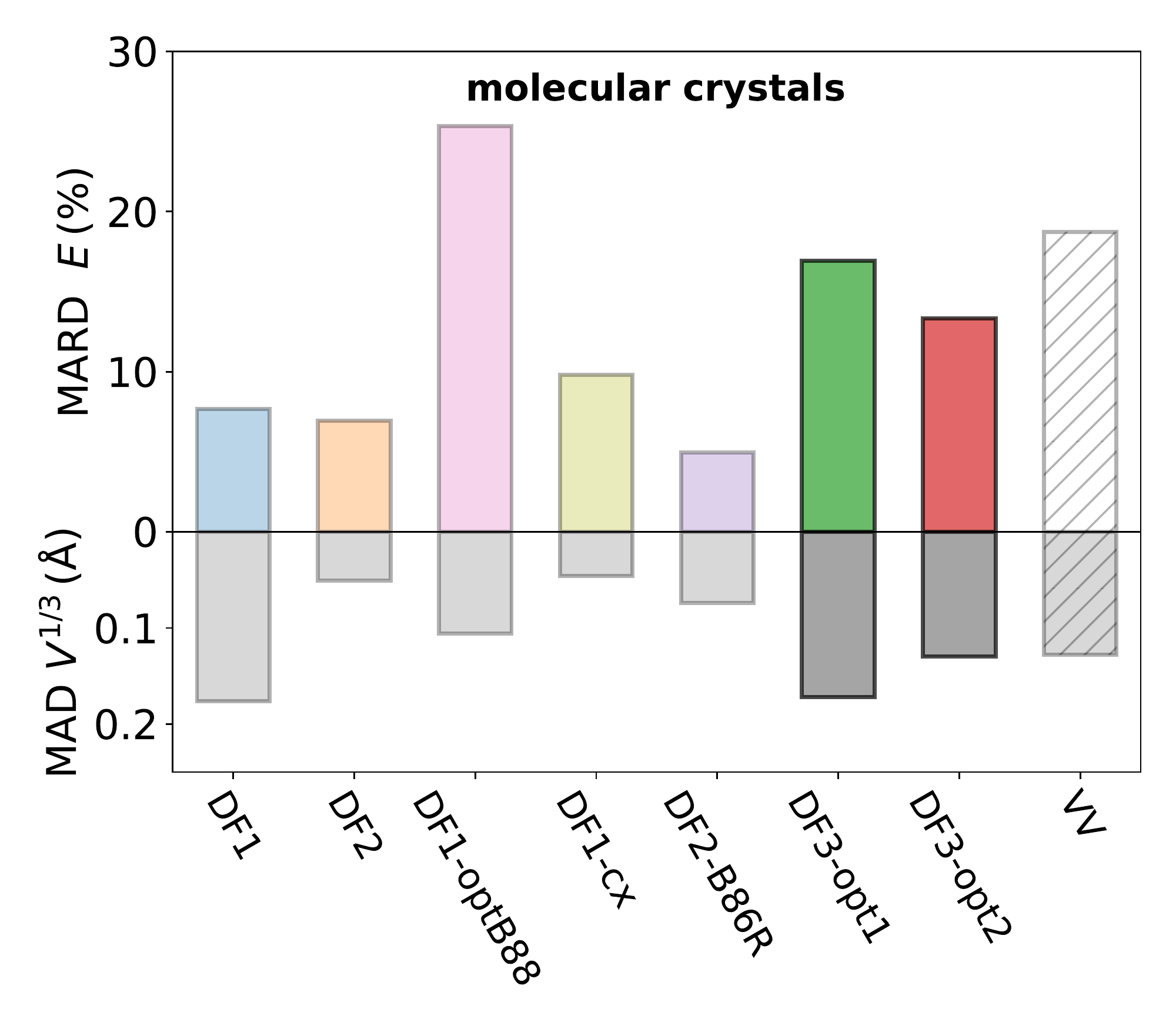}\\[-2ex]
\caption{\label{fig:molecular_crystals}MARD of the cohesive energy per
monomer (with respect to separation into molecules) and MAD of the
third-root volume for the X23 dataset of molecular crystals.}
\end{figure}

\begin{figure*}
\includegraphics[width=\textwidth]{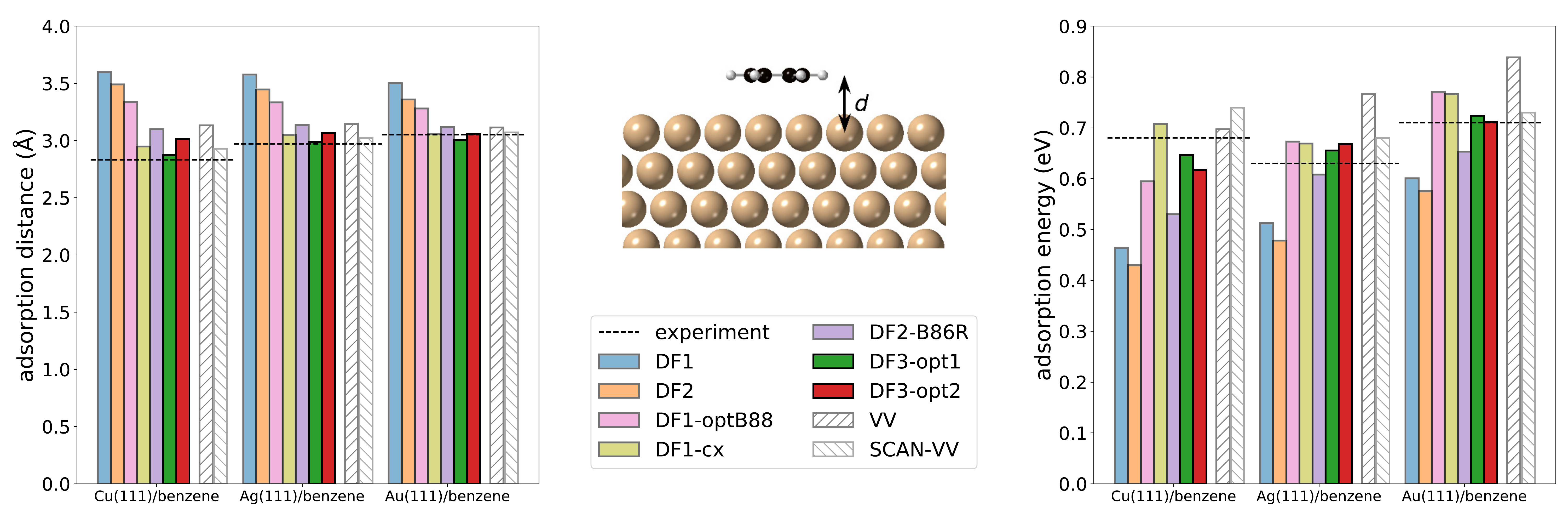}\\[-2ex]
\caption{\label{surface_adsorption} {\bf(left)} Adsorption distance
(\AA) and {\bf(right)} adsorption energy (eV) for benzene on Cu, Ag, and
Au. Experimental reference values are taken from
Ref.~\onlinecite{Maass_2018:binding_energies}. The inset shows the
adsorption geometry.}
\end{figure*}

\subsection{Molecular Crystals}

An important benchmark for all van der Waals functionals are molecular
crystals.\cite{Berland_2015:van_waals, Berland_2011:van_waals,
Rangel_2016:structural_excited-state,Tran_2019:nonlocal_vdW,
Carter_2014:benchmarking_calculated,
Otero-de-la-Roza_2012:benchmark_non-covalent, Peng_2019:van_waals,
Reilly_2013:understanding_role} We have calculated the optimized volume
and the cohesive energy per monomer (with respect to separation into
molecules) for the X23 dataset of molecular
crystals.\cite{doi:10.1021/jp501237c} Results are depicted in
Fig.~\ref{fig:molecular_crystals}; a summary of statistical data is
available in Appendix~\ref{sec:statistics} and detailed results for each
crystal are provided in the Supporting Information. Looking at the
third-root volume (as an average representative of lattice constants),
we see that our vdW-DF3 functionals are comparable to VV and DF1, but
they are less accurate than the other functionals. For the cohesive
energy, our functionals are noticeably more accurate than DF1-optB88 and
somewhat better that VV. Nonetheless, we do not recommend our two
vdW-DF3 functional variants for molecular crystals, as better options
are available. The underestimation of vdW-DF3 volumes can be linked to
the shape of $F_{\rm x}(s)$ and $d F_{\rm x}(s)/ds$ as shown in
Ref.~\onlinecite{Peng_2019:van_waals} and is the result of a conscious
trade-off we made for our $\kappa$ values in Eqs.~(\ref{eq:Fx-b86r}) and
(\ref{eq:Fx-optb88}), see the discussion in Section~\ref{sec:balance}.
For the underestimation of cohesive energies, we speculate that the
enhancement factor also plays an important role, but other well-known
effects such as the delocalization error may also contribute---again,
see Section~\ref{sec:balance} for more discussion. Although molecular
crystals are bound by much the same interactions as present in our
fitting set, the relative spatial configurations/orientations between
molecules in the X23 can differ from those present in the S22$\times$5
and molecular crystals are denser in the sense that the average shortest
distances between molecules in S22 is 13\% larger than in the X23 set.
In addition, many-body dispersion effects play a significant role in the
molecular crystals of the X23 set, while they are rather small in
molecular dimers.\cite{doi:10.1021/ar500144s}

\subsection{Benzene Adsorption on Cu/Ag/Au (111)}

Finally, we benchmark our new functionals also against molecular
adsorption on coinage metals, which are challenging
systems.\cite{Berland_2014:van_waals} In particular, we study the
adsorption of benzene on the (111) surface of Cu, Ag, and Au. A summary
of statistical data is available in Appendix~\ref{sec:statistics}. In
Fig.~\ref{surface_adsorption} we show the benzene adsorption distance
from the surface and its adsorption energy and we see that the original
DF1 and DF2 functionals significantly overestimate the binding
separations, resulting in dramatic consequences for surface
corrugation.\cite{Lee_2012:benchmarking_van} This figure also shows
nicely the progress that has been made within the vdW-DF family, with
DF3-opt1 providing distances that are almost spot-on the reference data
and very accurate energies, closely followed by DF3-opt2. This good
performance could be related to the excellence performance for
dispersion-dominated systems in Fig.~\ref{S66x8} for larger-than-binding
separations. The adsorbed molecule interacts with the surface through
dispersion forces not only with its footprint directly vertically
underneath at typical binding distances, but also with the horizontally
surrounding surface at larger-than-binding separations---and this is
where our new functionals excel. The importance of such horizontal
interactions beyond-binding separations has been demonstrated in
Ref.~\onlinecite{Berland_2013:analysis_of_vdW}. This aspect is also
intimately linked to---and paralleled by---our improved performance for
layered systems.

\section{Balancing Competing Interests --- What can be Expected
from the vdW-DF Framework?}\label{sec:balance}

The results in the previous sections showed that our new functionals
vdW-DF3-opt1 and vdW-DF3-opt2 perform very well. The main advancement is
the greatly increased performance for dispersion-dominated molecular
dimers, especially at larger-than-binding separations, see
Fig.~\ref{S66x8}. Although we also see improved and generally good
performance for many other systems, we would like to point out that
performance---although still good and comparable to SCAN+VV and VV and
better than vdW-DF1 and PBE+D3---is somewhat less accurate for
hydrogen-bonded systems at their equilibrium separation.

We have noticed this trend early on and investigated measures to also
achieve highly accurate performance for hydrogen-bonded systems at the
equilibrium separation. These systems are very much controlled by the
choice of exchange and we have investigated further parameterized
versions of Eq.~(\ref{eq:Fx-b86r}), where changing $c$ in conjunction
with $\kappa$ would, in fact, lead exactly to the desired improvement
and we see good performance for hydrogen-bonded dimers around the
equilibrium separation and molecular crystals. However, through this
higher dimensional parameter search (and other avenues we have
investigated) we learned an important lesson: With our new development,
the overall vdW-DF framework is coming to its performance limits.
Although possible new $h$-functions provide additional degrees of
freedom that allow for improvements of many aspects of particular
systems, we now see that further improvements are only possible to the
detriment of other areas. In our case, improving the hydrogen-bonded
systems at binding separation would lead to a decrease in accuracy for
dispersion bound dimers, layered systems, and surface adsorption.

\begin{figure*}
\includegraphics[width=\textwidth]{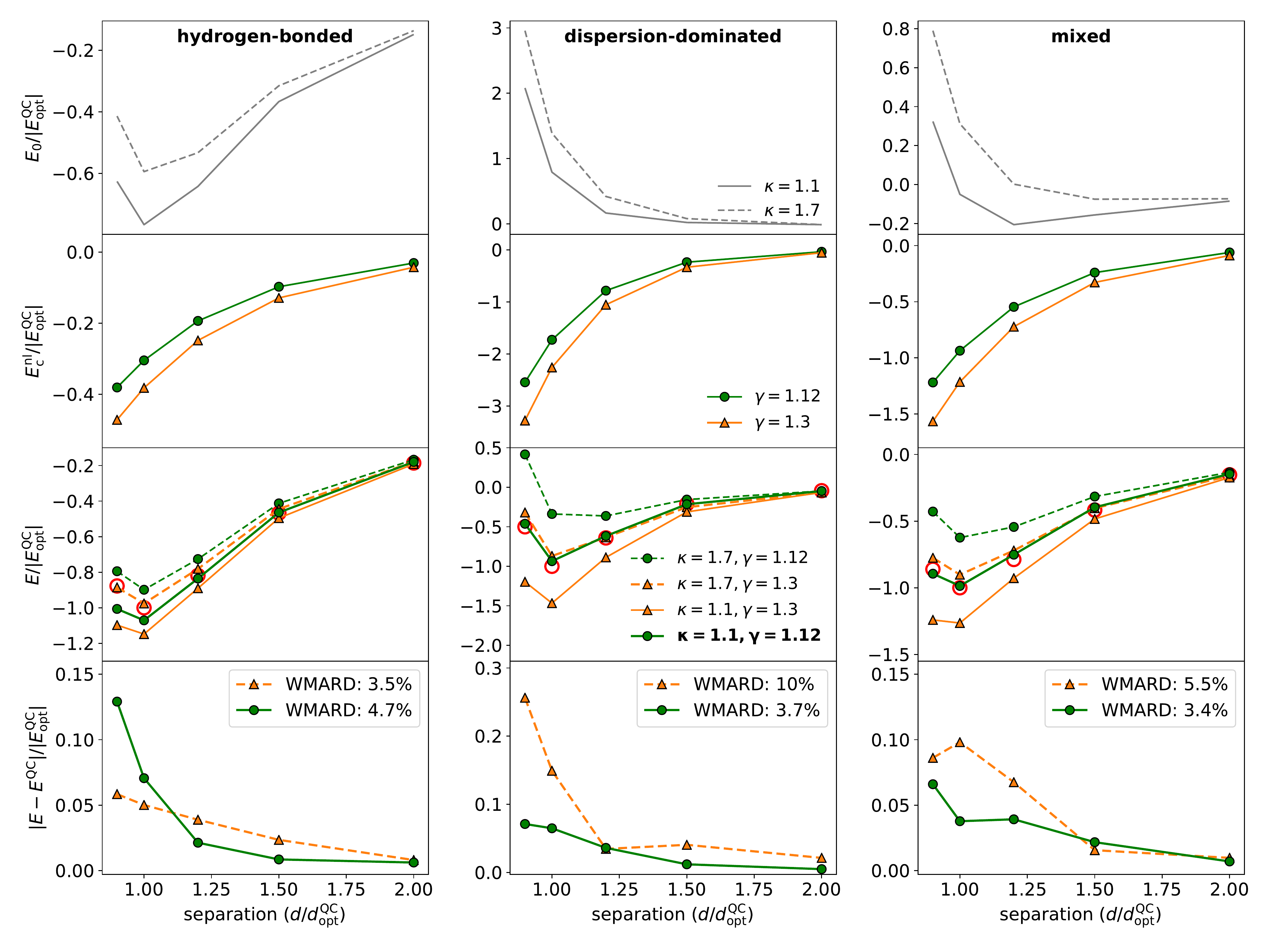}\\[-3ex]
\caption{\label{fig:E0-Ecnl}Split-up of the total energy into its
non-local contribution (first row) and the rest (second row), i.e.\ $E =
E_{\rm c}^{\rm nl} + E_0$ (third row), averaged over hydrogen-bonded,
dispersion-dominated, and mixed dimers of the S22$\times$5 dataset. All
energies are reported relative to the QC energy at the optimal
separation $E_{\rm opt}^{\rm QC}$ and have been averaged over all dimers
in that set. In the third row, QC reference data are indicated by red
circles. Curves in third row combine the style from the top panel and
color from the middle panel. In the fourth row, we show ${\rm
WMARD_{sep}}$ from Eq.~(\ref{WMARD_sep}) for DF3-opt1 with
$\kappa=1.1$/$\gamma=1.12$ and the functional with
$\kappa=1.7$/$\gamma=1.3$; the legend gives the mean over all five
distances.}
\end{figure*}

We show in Fig.~\ref{fig:E0-Ecnl} how the balancing of competing
interests plays out for the case of hydrogen-bonded molecular dimers
vs.\ dispersion-dominated molecular dimers in Fig.~\ref{S66x8}. In
particular, we study the split-up of the total energy into its non-local
contribution $E_{\rm c}^{\rm nl}$ and the rest $E_0$, i.e.\ $E = E_{\rm
c}^{\rm nl} + E_0$. Figure~\ref{fig:E0-Ecnl} shows this split-up as a
function of our parameters $\kappa$ and $\gamma$. Our choice for
DF3-opt1 was $\kappa=1.1$/$\gamma=1.12$, leading to very good
performance for dispersion-dominated systems and somewhat less accuracy
for hydrogen-bonded systems around and below the binding separation.
However, we see that a choice of $\kappa=1.7$/$\gamma=1.3$ would have
reversed those roles. In the Supporting Information we show the shape
of $F_{\rm x}(s)$ and $dF_{\rm x}(s)/ds$ for the $\kappa=1.7$
parameterization in comparison to the ones in Fig.~\ref{fig:Fx}---one
can see that the derivative starts to be noticeably larger for
$\kappa=1.7$ compared to $\kappa=1.1$ for $s>0.5$. The choice of
optimizing WMARD of the S$22\times5$ set as described in
Sec.~\ref{sec:optimization_scheme} resulted in dispersion-bonded systems
being favored at the expensive of hydrogen-bonded systems. This is
because the dispersion part is far more sensitive to the parameter
choice for the dispersion-bonded systems, as can be seen in
Fig.~\ref{fig:E0-Ecnl}. The hydrogen-bonded part of WMARD is also
significantly smaller in magnitude compared to the dispersion-dominated
part. Our choice is in line with the fact that dispersion-bonded
systems were the original target of the vdW-DF development and because
of their impact on a large class of relevant problems in surface
adsorption and layered structures. We speculate that our choice of
$\kappa$ in vdW-DF3-opt1 and vdW-DF3-opt2 is also responsible for the
lower accuracy in molecular crystals. We note in this context that the
overall MAD of the S$22\times5$ set is in fact smaller for
$\kappa=1.7$/$\gamma=1.3$, i.e.\ 11.0~meV compared to 13.2 meV for
vdW-DF3-opt1, but the latter has a considerable smaller MARD, i.e.\
8.3\% compared to 18.6\% for $\kappa=1.7$/$\gamma=1.3$. Finally, the
fourth column of Fig.~\ref{fig:E0-Ecnl} highlights the reasons for our
choice of $\kappa$ and $\gamma$: While better performance in terms of
WMARD could have been achieved for hydrogen-bonded systems with a choice
of $\kappa=1.7$/$\gamma=1.3$, that same choice would result in
noticeably reduced performance for dispersion-dominated and mixed
systems (see legend of fourth row in Fig.~\ref{fig:E0-Ecnl}).

To describe van der Waals interactions and non-covalent interactions
beyond van der Waals interactions---in particular halogen- and
hydrogen-bonded systems---with high accuracy at the same time is a
long-standing problem within DFT and has been linked to the
delocalization error resulting from incomplete self-interaction
correction.\cite{Cohen792, doi:10.1021/ct500899h,
doi:10.1021/acs.jctc.9b00550} This error is present in vdW-DF through
the choice of (semi)local exchange and correlation and it is conceivable
that the trade-off and reduced accuracy in certain systems are related
to this error. Recently, a BEEF-vdW+$U$ method has been developed that
tries to counteract the delocalization error through a Hubbard $U$ type
correction and does indeed show improved results, albeit for strongly
correlated systems where the error is more prominent.\cite{Li_2019} In
this context, we would also like to point out that
PBE+D3\cite{doi:10.1063/1.3382344} has good performance for molecular
dimers\cite{doi:10.1002/cphc.201100826} and at the same time very good
performance for molecular crystals, see
Ref.~\onlinecite{doi:10.1021/jp501237c} and the Supporting Information
for a comparison. This points towards a particular quality of PBE+D3
that is capable to overcome the trade-off inherent in vdW-DF.
Reference~\onlinecite{C9CS00060G} shows that the good MARD for cohesive
energies of PBE+D3 for molecular crystals almost triples to 16\%---quite
comparable to vdW-DF3---when higher multipole interactions are
neglected, giving important insight that may help future vdW-DF
developments. PBE+D3 does, however, have difficulties with surface
adsorption and layered structures.\cite{Tran_2019:nonlocal_vdW,
Reckien2014}

Through the various improvements of the vdW-DF framework over the years
we have reached a point where the performance of the original vdW-DF
framework has been pushed to its limit and the fundamental design
choices are now becoming the bottleneck. We see three exciting ways
forward: (i) the inclusion of some fraction of exact exchange in the
pairing with the vdW-DF non-local correlation. This is already ongoing
work and shows promise.\cite{doi:10.1063/1.4986522,
doi:10.1063/1.5012870} This approach has the potential to overcome the
current trade-off we encountered between hydrogen-bonded systems and
dispersion-dominated systems---in particular with the additional degree
of freedom in $h(y)$---and also provides a systematic improvement of the
delocalization error.\cite{doi:10.1021/acs.jctc.9b00550} (ii) New
functionals within the vdW-DF family could be developed for specific
applications, rebalancing our choice. Such functionals would be somewhat
limited in scope, but can show very good accuracy for the situation they
have been designed for. Applications of particular interest may be
adsorption systems, molecular crystals, or transition-state theory
calculations. (iii) Alternatively, it is possible to fundamentally
change the vdW-DF framework and deviate from its original design
philosophy. We see this as the only option to achieve high accuracy for
all systems at the same time and thus truly generate a general-purpose
functional. So, where would one even start thinking about such a
fundamental change? Below Eq.~(\ref{equ:omega}) we point out that vdW-DF
uses only a single length-scale to parameterize its plasmon-dispersion
model. Already in the 2004 paper we see that this is an approximation
made for convenience,\cite{Dion_2004:van_waals} and the introduction of
a second length scale would be beneficial. It is, in fact, surprising
that the vdW-DF framework captures such a diverse group of vastly
different types of materials so reasonably well. Another possible
direction could be to update the rather simple vdW-DF plasmon-dispersion
model altogether, maybe along the lines of the VV functionals, from
which much can be learned. Finally, it is conceivable that a focus on
different physical constraints leads to a more accurate form for $S$ in
Eq.~(\ref{eq:S1}) or maybe $S$ could be approximated through better
models for the response function. However, common to several of these
directions would be that they fundamentally change the vdW-DF framework
and design philosophy to such a point that they present completely new
directions and thus would likely no longer carry the original vdW-DF
name.

\section{Conclusions}

We have presented the next-generation non-local van der Waals density
functional vdW-DF3. It is entirely built within the design guidelines of
the original vdW-DF, but takes advantage of a newly discovered degree of
freedom within the framework to significantly improve performance, in
particular for beyond-binding separations. At the same time, we show
that---by observing the vdW-DF constraints and building on lessons
learned in successive developments---vdW-DF3 can retain the same wide
transferability as earlier variants. This finding is based on
benchmarking on a wide array of systems, in which we also compare with
earlier van der Waals functionals, allowing us to document successive
improvements. While we find generally good performance of vdW-DF3 for
many systems, the most striking improvement is found for
dispersion-dominated systems beyond binding separation. Our analysis
also indicates that, with recent developments in general and vdW-DF3 in
particular, the vdW-DF framework is operating close to its limits in
terms of overall accuracy. This is also evident through the similarity
of the DF3-opt2 parametrization of vdW-DF3 and the DF2-B86R functional.
However, as the vdW-DF3 design is more flexible than its predecessors,
it opens the door for functionals tailored to more specific classes of
systems, which will likely cause some worsening in other areas.
Finally, we provide an outlook for research directions that could
overcome the fundamental bottlenecks of the vdW-DF framework and lead to
further improvements for even broader classes of systems.

\section*{Acknowledgement}

This work was supported by the U.S.\ National Science Foundation Grant
No.\ DMR--1712425. KB also acknowledges auxiliary funding from the
Research Council of Norway No.\ 302362. We also thank Tonatiuh Rangel
for providing some initial molecular crystal structures. The majority of
calculations were performed on the WFU DEAC cluster, with some parts
utilizing resources provided by \textsc{uninett} Sigma2 (National
Infrastructure for High Performance Computing in Norway). We also
gratefully acknowledge Per Hyldgaard for providing input files for the
G2--1 and G2RC sets.

\clearpage
\begin{appendix}

\onecolumngrid\section{Statistical Data}\label{sec:statistics}

\vspace*{-3ex}

\begin{table*}[h!]\footnotesize
\caption{\label{table:s22x5-all}Comparison of mean deviation (MD), mean
absolute deviation (MAD), mean absolute relative deviation (MARD) from
Eq.~(\ref{MARD}) and weighted mean absolute relative deviation (WMARD)
from Eq.~(\ref{WMARD}) for the interaction energies of the S22$\times$5
set of molecular dimers for all separations. Deviations are reported
with respect to quantum chemistry calculations at the CCSD(T) level from
Ref.~\onlinecite{Pavel_2010:comparative_study_s22x5}.}
\begin{tabular*}{\textwidth}{@{}l@{\extracolsep{\fill}}rrrrrrrrr@{}}\hline\hline
Complex            &      DF1 &      DF2 &  DF1-optb88 &      DF1-cx &  DF2-B86R &  DF3-opt1 & DF3-opt2 &     VV &  SCAN-VV \\\hline
\multicolumn{10}{@{}l}{\it Hydrogen bonded complexes (7)}\\
MD [meV]           &   54.05  &    31.27 &        0.84 &       10.53 &      9.09 &  $-$24.26 & $-$12.01 &$-$18.84 & $-$21.11\\
MAD [meV]          &   60.71  &    34.48 &        9.40 &       17.55 &     11.00 &     28.42 &    14.29 &  20.02 &    28.75 \\
MARD [\%]          &   14.36  &     7.01 &        3.17 &        6.28 &      3.02 &      6.25 &     3.81 &   5.11 &     7.41 \\
WMARD [\%]         &   11.04  &     5.63 &        2.17 &        4.00 &      2.08 &      4.72 &     2.64 &   3.83 &     4.71 \\[2ex]
\multicolumn{10}{@{}l}{\it Complexes with predominant dispersion contribution (8)}\\
MD [meV]           &    29.40 &    30.84 &     $-$6.23 &        2.99 &     21.57 &      4.67 &     3.99 &  12.75 &    14.14 \\
MAD [meV]          &    58.14 &    36.82 &       14.59 &       26.04 &     21.77 &      6.25 &     5.78 &  15.23 &    15.12 \\
MARD [\%]          &   135.84 &    74.36 &       41.29 &       67.94 &     42.88 &     12.44 &    13.98 &  33.09 &    67.42 \\
WMARD [\%]         &    32.78 &    17.88 &        9.92 &       18.82 &     13.23 &      3.67 &     3.69 &   7.32 &    10.11 \\[2ex]
\multicolumn{10}{@{}l}{\it Mixed complexes (7)}\\
MD [meV]           &    17.40 &    19.25 &        1.96 &        5.58 &     16.00 &      3.15 &     3.89 &   6.60 &     6.24 \\
MAD [meV]          &    28.61 &    19.91 &        6.67 &       15.23 &     16.07 &      5.91 &     5.54 &   7.73 &    10.34 \\
MARD [\%]          &    26.29 &    15.10 &        8.35 &       17.08 &     12.24 &      5.71 &     6.47 &   7.12 &    12.55 \\
WMARD [\%]         &    16.97 &    11.06 &        4.30 &        9.24 &      8.94 &      3.62 &     3.51 &   4.84 &     6.56 \\[2ex]
\multicolumn{10}{@{}l}{\it Average over all separation for all complexes (22)}\\
MD [meV]           &    33.43 &    27.29 &     $-$1.37 &        6.21 &     15.83 &   $-$5.02 &  $-$1.13 &   0.74 &     0.41 \\
MAD [meV]          &    49.56 &    30.70 &       10.42 &       19.90 &     16.53 &     13.19 &     8.41 &  14.37 &    17.94 \\
MARD [\%]          &    62.33 &    34.08 &       18.68 &       32.14 &     20.45 &      8.33 &     8.35 &  15.92 &    30.87 \\
WMARD [\%]         &    20.83 &    11.81 &        5.67 &       11.06 &      8.32 &      3.99 &     3.30 &   5.42 &     7.26 \\\hline\hline
\end{tabular*}
\end{table*}

\begin{table*}[h!]\footnotesize
\caption{\label{table:s66x8-all}Comparison of various statistical
measures for the interaction energies of the S66$\times$8 set of
molecular dimers for all separations. See caption of
Table~\ref{table:s22x5-all} for more details. Reference data taken from
Ref.~\onlinecite{Rezac_2011:s66_well-balanced}.}
\begin{tabular*}{\textwidth}{@{}l@{\extracolsep{\fill}}rrrrrrrrr@{}}\hline\hline
System                  &     DF1  &       DF2 & DF1-optb88 &    DF1-cx  &   DF2-B86R &  DF3-op1 &DF3-opt2 &      VV & SCAN-VV \\\hline
\multicolumn{10}{@{}l}{\it Hydrogen bonded complexes (23)}\\
MD [meV]                &    32.88 &     12.81 &    $-$3.01 &       6.47 &       4.32 & $-$21.20 &$-$14.89 &$-$17.43 &$-$18.08 \\
MAD [meV]               &    39.45 &     17.44 &       7.29 &      13.87 &       9.00 &    22.19 &   15.27 &   17.89 &   20.65 \\
MARD [\%]               &    12.92 &      4.85 &       3.36 &       6.40 &       3.13 &     6.46 &    5.47 &    5.74 &    6.27 \\
WMARD  [\%]             &    10.00 &      3.94 &       2.22 &       4.12 &       2.53 &     5.79 &    4.60 &    4.98 &    5.04 \\[2ex]
\multicolumn{10}{@{}l}{\it Complexes with predominant dispersion contribution (23)}\\
MD [meV]                &     7.65 &      7.52 &   $-$17.68 &    $-$2.30 &      14.19 &  $-$4.13 &   -6.41 & $-$0.57 &    10.09 \\
MAD [meV]               &    38.54 &     20.27 &      18.02 &      21.08 &      15.04 &     5.91 &    7.38 &    8.05 &    10.87 \\
MARD [\%]               &    69.98 &     29.83 &      35.14 &      42.89 &      17.45 &    10.59 &   12.00 &   14.74 &    16.14 \\
WMARD [\%]              &    26.31 &     13.52 &      13.35 &      14.89 &       9.73 &     4.73 &    6.18 &    5.74 &     7.85 \\[2ex]
\multicolumn{10}{@{}l}{\it Others (20)}\\
MD [meV]                &    14.69 &     12.75 &    $-$2.65 &       4.99 &      13.68 &  $-$1.49 & $-$1.68 &    1.79 &     1.68 \\
MAD [meV]               &    28.00 &     16.14 &       7.76 &      15.58 &      13.93 &     5.57 &    5.55 &    6.31 &     9.62 \\
MARD [\%]               &    28.62 &     14.17 &      10.62 &      18.80 &      11.74 &     4.87 &    6.24 &    5.91 &     9.70 \\
WMARD [\%]              &    17.94 &     10.12 &       5.19 &      10.25 &       8.88 &     3.54 &    3.74 &    4.12 &     6.07 \\[2ex]
\multicolumn{10}{@{}l}{\it Average over all separation for all complexes (66)}\\
MD [meV]                &   18.57 &     10.95 &    $-$8.02 &       2.97 &      10.59 &  $-$9.28 & $-$7.93 & $-$5.73 &  $-$2.28 \\
MAD [meV]               &   35.66 &     18.03 &      11.17 &      16.90 &      12.60 &    11.48 &    9.58 &   10.95 &    13.90 \\
MARD [\%]               &   37.56 &     16.38 &      16.64 &      22.87 &      10.73 &     7.42 &    7.98 &    8.93 &    10.75 \\
WMARD [\%]              &   18.09 &      9.15 &       7.00 &       9.73 &       6.96 &     4.74 &    4.89 &    4.98 &     6.33 \\\hline\hline
\end{tabular*}
\end{table*}

\clearpage

\begin{table*}[h!]\footnotesize
\caption{\label{table:G2-1}Comparison of various statistical measures
for the atomization energies of molecules from the G2--1
set\cite{doi:10.1063/1.460205, doi:10.1063/1.473182} and reaction
energies from the G2RC set.\cite{doi:10.1063/1.473182} Reference data
taken from Refs.~\onlinecite{Grimme_website, doi:10.1063/1.1926272}.
For the G2RC set we also report the root-mean-square deviation (RMSD)
for easier comparison with Ref.~\onlinecite{Grimme_website}. Further
details are available in the Supporting Information.}
\begin{tabular*}{\textwidth}{@{}l@{\extracolsep{\fill}}rrrrrrrrrr@{}}\hline\hline
System      &     PBE  &   PBEsol &    DF1  &      DF2 &DF1-optB88&   DF1-cx & DF2-B86R & DF3-opt1 & DF3-opt2 &       VV \\\hline
\multicolumn{9}{@{}l}{\it G2--1}\\
MD [eV]     &  $-$0.39 &  $-$0.76 & $-$0.12 &  $-$0.03 &  $-$0.37 &  $-$0.48 &  $-$0.47 &  $-$0.65 &  $-$0.52 &  $-$0.23 \\
MAD [eV]    &     0.49 &     0.79 &    0.24 &     0.26 &     0.40 &     0.50 &     0.49 &     0.66 &     0.54 &     0.38 \\
MARD [\%]   &     9.04 &    12.47 &    5.15 &     5.16 &     7.90 &     8.70 &     8.64 &    10.85 &     9.19 &     7.72 \\[2ex]
\multicolumn{9}{@{}l}{\it G2RC}\\
MD [eV]     &     0.04 &  $-$0.09 &    0.29 &     0.33 &     0.15 &     0.10 &     0.11 &     0.02 &     0.08 &     0.11 \\
MAD [eV]    &     0.30 &     0.45 &    0.35 &     0.43 &     0.28 &     0.33 &     0.29 &     0.34 &     0.30 &     0.25 \\
RMSD [eV]   &     0.36 &     0.53 &    0.43 &     0.51 &     0.36 &     0.42 &     0.37 &     0.43 &     0.38 &     0.33 \\\hline\hline
\end{tabular*}
\end{table*}

\clearpage

\begin{table*}[h!]\footnotesize
\caption{\label{table:solids_LC} Lattice constants [\AA] and atomization
energies [eV] for selected solids. Zero-point corrected experimental
lattice constants and atomization energies are taken from
Ref.~\onlinecite{Klimes_2011:van_waals,
Csonka_2009:assessing_performance, Peng_2016:versatile_vdW_SCAN-rVV10}
and references therein; SCAN-VV data taken from
Ref.~\onlinecite{Peng_2016:versatile_vdW_SCAN-rVV10}.}
\begin{tabular*}{\textwidth}{@{}l@{\extracolsep{\fill}}rrrrrrrrrrrr@{}}\hline\hline
System    &       expt. &  PBE  &  PBEsol&   DF1 &  DF2 &  DF1-optB88 &  DF1-cx &  DF2-B86R &  DF3-opt1 &  DF3-opt2 &   VV &  SCAN-VV\\\hline
\multicolumn{13}{@{}l}{\it Lattice constants}\\
Cu        &        3.60 &   3.63 &    3.56 &    3.69 &    3.75 &        3.62 &    3.57 &      3.59 &      3.57 &      3.59 & 3.65 &     3.54 \\
Ag        &        4.06 &   4.15 &    4.05 &    4.25 &    4.33 &        4.14 &    4.07 &      4.11 &      4.08 &      4.10 & 4.17 &     4.06 \\
Pd        &        3.88 &   3.95 &    3.88 &    4.01 &    4.09 &        3.94 &    3.89 &      3.92 &      3.90 &      3.91 & 3.98 &     3.88 \\
Rh        &        3.79 &   3.83 &    3.78 &    3.88 &    3.95 &        3.84 &    3.79 &      3.81 &      3.80 &      3.81 & 3.87 &     3.77 \\
Na        &        4.21 &   4.20 &    4.17 &    4.21 &    4.14 &        4.15 &    4.24 &      4.17 &      4.14 &      4.15 & 4.13 &     4.15 \\
K         &        5.21 &   5.28 &    5.21 &    5.31 &    5.20 &        5.21 &    5.32 &      5.24 &      5.22 &      5.21 & 5.14 &     5.23 \\
Rb        &        5.58 &   5.67 &    5.57 &    5.60 &    5.51 &        5.49 &    5.59 &      5.53 &      5.48 &      5.51 & 5.46 &     5.59 \\
Cs        &        6.04 &   6.16 &    6.01 &    6.00 &    5.93 &        5.83 &    5.90 &      5.89 &      5.80 &      5.86 & 5.84 &     6.05 \\
Ca        &        5.55 &   5.53 &    5.46 &    5.54 &    5.48 &        5.44 &    5.46 &      5.46 &      5.41 &      5.44 & 5.46 &     5.52 \\
Sr        &        6.05 &   6.03 &    5.92 &    6.07 &    6.02 &        5.92 &    5.93 &      5.94 &      5.88 &      5.92 & 5.93 &     6.04 \\
Ba        &        5.00 &   5.02 &    4.88 &    5.07 &    5.05 &        4.91 &    4.87 &      4.92 &      4.85 &      4.90 & 4.92 &     4.98 \\
Al        &        4.02 &   4.04 &    4.01 &    4.09 &    4.09 &        4.06 &    4.03 &      4.04 &      4.03 &      4.04 & 4.03 &     4.00 \\
LiF       &        3.96 &   4.07 &    4.01 &    4.11 &    4.08 &        4.03 &    4.06 &      4.04 &      4.00 &      4.02 & 4.03 &     3.94 \\
LiCl      &        5.06 &   5.15 &    5.06 &    5.22 &    5.21 &        5.12 &    5.11 &      5.11 &      5.06 &      5.09 & 5.11 &     5.04 \\
NaF       &        4.58 &   4.72 &    4.65 &    4.76 &    4.70 &        4.66 &    4.71 &      4.67 &      4.62 &      4.65 & 4.65 &     4.53 \\
NaCl      &        5.57 &   5.70 &    5.61 &    5.75 &    5.70 &        5.63 &    5.67 &      5.64 &      5.58 &      5.61 & 5.61 &     5.51 \\
MgO       &        4.18 &   4.26 &    4.22 &    4.28 &    4.29 &        4.23 &    4.23 &      4.23 &      4.21 &      4.23 & 4.25 &     4.17 \\
C         &        3.54 &   3.57 &    3.56 &    3.59 &    3.61 &        3.58 &    3.57 &      3.57 &      3.57 &      3.57 & 3.59 &     3.55 \\
SiC       &        4.34 &   4.38 &    4.36 &    4.40 &    4.43 &        4.38 &    4.37 &      4.38 &      4.37 &      4.37 & 4.40 &     4.35 \\
Si        &        5.42 &   5.47 &    5.44 &    5.51 &    5.55 &        5.48 &    5.44 &      5.46 &      5.45 &      5.46 & 5.50 &     5.42 \\
Ge        &        5.64 &   5.76 &    5.67 &    5.84 &    5.94 &        5.73 &    5.67 &      5.71 &      5.68 &      5.70 & 5.80 &     5.63 \\
GaAs      &        5.64 &   5.75 &    5.67 &    5.84 &    5.93 &        5.74 &    5.68 &      5.72 &      5.69 &      5.71 & 5.79 &     5.64 \\[2ex]
MD [\AA]  &  ---        &   0.06 &    0.00 &    0.10 &    0.09 &        0.01 &    0.01 &      0.01 &   $-$0.02 &  $-$0.002 & 0.02 &  $-$0.01 \\
MAD [\AA] &  ---        &   0.07 &    0.04 &    0.10 &    0.13 &        0.07 &    0.06 &      0.06 &      0.06 &      0.06 & 0.09 &     0.02 \\
MARD [\%] &  ---        &   1.44 &    0.73 &    2.20 &    2.75 &        1.47 &    1.14 &      1.17 &      1.10 &      1.12 & 1.80 &     0.43 \\[2ex]
\multicolumn{13}{@{}l}{\it Atomization energies}\\
Cu        &        3.52 &   3.53 &    4.10 &    3.07 &    2.89 &        3.59 &    3.90 &      3.70 &      3.95 &      3.79 & 3.72 &     4.04 \\
Ag        &        2.98 &   2.52 &    3.09 &    2.29 &    2.15 &        2.76 &    2.98 &      2.79 &      3.07 &      2.88 & 2.89 &     3.08 \\
Pd        &        3.94 &   3.74 &    4.46 &    3.36 &    3.16 &        4.00 &    4.32 &      4.09 &      4.39 &      4.19 & 4.13 &     4.59 \\
Rh        &        5.78 &   5.74 &    6.68 &    4.99 &    4.69 &        6.06 &    6.72 &      6.26 &      6.73 &      6.44 & 5.92 &     5.60 \\
Na        &        1.12 &   1.05 &    1.12 &    0.96 &    0.96 &        0.99 &    1.20 &      1.01 &      1.06 &      1.04 & 1.07 &     1.14 \\
K         &        0.94 &   0.86 &    0.92 &    0.82 &    0.73 &        0.85 &    0.90 &      0.84 &      0.89 &      0.86 & 0.94 &     0.91 \\
Rb        &        0.86 &   0.77 &    0.83 &    0.77 &    0.66 &        0.78 &    0.81 &      0.76 &      0.82 &      0.79 & 0.87 &     0.82 \\
Cs        &        0.81 &   0.71 &    0.78 &    0.79 &    0.64 &        0.76 &    0.77 &      0.72 &      0.79 &      0.75 & 0.87 &     0.75 \\
Ca        &        1.86 &   1.91 &    2.10 &    1.66 &    1.40 &        1.86 &    2.04 &      1.87 &      2.02 &      1.91 & 2.00 &     2.17 \\
Sr        &        1.73 &   1.61 &    1.81 &    1.42 &    1.12 &        1.61 &    1.78 &      1.60 &      1.76 &      1.63 & 1.73 &     1.91 \\
Ba        &        1.91 &   1.88 &    2.13 &    1.79 &    1.49 &        1.99 &    2.14 &      1.95 &      2.13 &      1.99 & 2.10 &     2.15 \\
Al        &        3.43 &   3.47 &    3.81 &    2.90 &    2.52 &        3.24 &    3.64 &      3.43 &      3.58 &      3.49 & 3.41 &     3.71 \\
LiF       &        4.46 &   4.39 &    4.49 &    4.44 &    4.57 &        4.56 &    4.44 &      4.49 &      4.59 &      4.55 & 4.55 &     4.49 \\
LiCl      &        3.59 &   3.36 &    3.49 &    3.42 &    3.44 &        3.54 &    3.51 &      3.49 &      3.59 &      3.54 & 3.53 &     3.58 \\
NaF       &        3.97 &   3.89 &    3.96 &    3.97 &    4.08 &        4.04 &    4.02 &      3.97 &      3.98 &      4.01 & 4.03 &     4.00 \\
NaCl      &        3.34 &   3.09 &    3.20 &    3.18 &    3.19 &        3.26 &    3.29 &      3.19 &      3.29 &      3.23 & 3.24 &     3.33 \\
MgO       &        5.20 &   5.11 &    5.38 &    4.95 &    4.93 &        5.06 &    5.27 &      5.20 &      5.26 &      5.26 & 5.26 &     5.34 \\
C         &        7.55 &   7.70 &    8.18 &    7.13 &    6.87 &        7.60 &    7.89 &      7.76 &      7.97 &      7.85 & 7.61 &     7.60 \\
SiC       &        6.48 &   6.38 &    6.79 &    5.95 &    5.72 &        6.38 &    6.61 &      6.48 &      6.68 &      6.56 & 6.34 &     6.55 \\
Si        &        4.68 &   4.51 &    4.86 &    4.19 &    4.00 &        4.55 &    4.75 &      4.62 &      4.80 &      4.69 & 4.57 &     4.82 \\
Ge        &        3.92 &   3.69 &    4.08 &    3.30 &    3.31 &        3.82 &    3.98 &      3.84 &      4.05 &      3.91 & 3.86 &     4.10 \\
GaAs      &        3.34 &   3.13 &    3.54 &    2.85 &    2.80 &        3.27 &    3.40 &      3.27 &      3.50 &      3.34 & 3.35 &     3.47 \\[2ex]
MD [eV]   & ---         &$-$0.11 &    0.20 & $-$0.33 & $-$0.46 &     $-$0.04 &    0.13 &  $-$0.004 &      0.16 &      0.06 & 0.03 &     0.12 \\
MAD [eV]  & ---         &   0.13 &    0.23 &    0.33 &    0.48 &        0.10 &    0.16 &      0.10 &      0.18 &      0.12 & 0.08 &     0.15 \\
MARD [\%] & ---         &   5.03 &    6.24 &   10.19 &   16.31 &        4.08 &    4.87 &      4.24 &      5.08 &      3.99 & 2.90 &     5.47 \\\hline\hline
\end{tabular*}
\end{table*}

\clearpage

\begin{table*}[h!]\footnotesize
\caption{\label{table:Layered_structure_LC} Layer separation [\AA] and
layer binding energies [meV/\AA$^2$] for selected layered structures.
Reference data for the separation is taken from experiment and for the
binding energy from RPA, as detailed in
Ref.~\onlinecite{Bjorkmann_2012:vdW_bonding_layered,
Bjorkman_2014:testing_several} and references therein. SCAN-VV data
taken from Ref.~\onlinecite{Peng_2016:versatile_vdW_SCAN-rVV10}.}
\begin{tabular*}{\textwidth}{@{}l@{\extracolsep{\fill}}rrrrrrrrrr@{}}\hline\hline
System    &   ref. &  DF1 &  DF2 &  DF1-optB88 &  DF1-cx &  DF2-B86R &  DF3-opt1 &  DF3-opt2 &   VV &  SCAN-VV \\\hline
\multicolumn{11}{@{}l}{\it Layer separation}\\
BN         &       3.35 &    3.54 &   3.47 &        3.28 &    3.21 &      3.26 &      3.24 &      3.23 &  3.29 &    3.24 \\
graphite   &       3.35 &    3.55 &   3.49 &        3.34 &    3.28 &      3.31 &      3.31 &      3.29 &  3.35 &    3.27 \\
HfS$_2$    &       5.84 &    5.95 &   6.00 &        5.82 &    5.75 &      5.78 &      5.74 &      5.77 &  5.87 &    5.79 \\
HfSe$_2$   &       6.16 &    6.46 &   6.53 &        6.30 &    6.23 &      6.27 &      6.22 &      6.26 &  6.39 &    6.14 \\
HfTe$_2$   &       6.65 &    7.25 &   7.27 &        6.78 &    6.55 &      6.67 &      6.59 &      6.63 &  6.81 &    6.69 \\
MoS$_2$    &       6.15 &    6.62 &   6.57 &        6.26 &    6.12 &      6.18 &      6.14 &      6.15 &  6.24 &    6.14 \\
MoSe$_2$   &       6.46 &    7.03 &   7.02 &        6.62 &    6.45 &      6.54 &      6.48 &      6.51 &  6.62 &    6.51 \\
PdTe$_2$   &       5.11 &    5.65 &   6.02 &        5.36 &    5.14 &      5.25 &      5.19 &      5.23 &  5.48 &    5.00 \\
WS$_2$     &       6.16 &    6.58 &   6.53 &        6.28 &    6.15 &      6.21 &      6.17 &      6.18 &  6.26 &    6.12 \\[2ex]
MD [\AA]   &        --- &    0.58 &   0.59 &        0.12 & $-$0.07 &      0.03 &   $-$0.03 &   $-$0.01 &  0.15 & $-$0.04 \\
MAD [\AA]  &        --- &    0.58 &   0.59 &        0.16 &    0.09 &      0.10 &      0.08 &      0.09 &  0.18 &    0.06 \\
MARD [\%]  &        --- &    6.81 &   7.26 &        1.96 &    1.30 &      1.31 &      1.13 &      1.29 &  2.35 &    1.22 \\[2ex]
\multicolumn{11}{@{}l}{\it Layer binding energy}\\
BN         &       14.4 &   19.02 &  18.30 &       25.33 &   23.80 &     21.17 &     20.17 &     22.04 & 24.73 &   18.45 \\
graphite   &       18.3 &   20.54 &  20.21 &       27.00 &   25.27 &     23.28 &     21.01 &     23.95 & 26.81 &   20.30 \\
HfS$_2$    &       16.1 &   15.08 &  16.33 &       21.51 &   20.20 &     19.16 &     17.90 &     20.25 & 23.42 &   15.85 \\
HfSe$_2$   &       17.0 &   15.68 &  16.22 &       21.38 &   20.48 &     19.08 &     18.08 &     19.94 & 23.76 &   16.10 \\
HfTe$_2$   &       18.6 &   15.72 &  16.04 &       22.71 &   24.55 &     21.97 &     21.37 &     23.04 & 26.93 &   17.99 \\
MoS$_2$    &       20.5 &   18.76 &  19.61 &       25.73 &   24.36 &     23.28 &     21.41 &     24.14 & 29.74 &   19.89 \\
MoSe$_2$   &       19.6 &   17.52 &  18.10 &       24.73 &   24.37 &     22.34 &     21.11 &     23.04 & 29.61 &   19.33 \\
PdTe$_2$   &       40.1 &   23.59 &  21.73 &       42.86 &   51.60 &     44.33 &     46.73 &     45.95 & 47.43 &   41.74 \\
 WS$_2$    &       20.2 &   18.07 &  18.84 &       25.83 &   24.23 &     23.17 &     21.42 &     23.97 & 29.97 &   23.38 \\[2ex]
MD [meV/\AA$^2$] &  --- & $-$2.31 &$-$2.16 &        5.81 &    5.43 &      3.67 &      2.71 &      4.61 &  8.62 &    0.91 \\
MAD [meV/\AA$^2$]&  --- &    3.84 &   3.50 &        5.81 &    5.43 &      3.67 &      2.71 &      4.61 &  8.62 &    1.50 \\
MARD [\%]        &  --- &   16.08 &  13.53 &       32.36 &   26.31 &     19.60 &     13.55 &     24.38 & 45.67 &    8.15 \\\hline\hline
\end{tabular*}
\end{table*}

\clearpage

\begin{table*}[h!]\footnotesize
\caption{\label{table:molecular_crystals}Comparison of various
statistical measures for the cohesive energy per monomer (with respect
to separation into molecules) [eV] and the third-root volume [\AA] for
the X23 set of molecular dimers. Reference data taken from
Ref.~\onlinecite{doi:10.1021/jp501237c}. Further details are available
in the Supporting Information.}
\begin{tabular*}{\textwidth}{@{}l@{\extracolsep{\fill}}rrrrrrrr@{}}\hline\hline
System            &      DF1 &      DF2 &  DF1-optb88 &      DF1-cx &  DF2-B86R &  DF3-opt1 & DF3-opt2 &     VV\\\hline
\multicolumn{9}{@{}l}{\bf Cohesive energy}\\
\multicolumn{9}{@{}l}{\it vdW-bonded systems (10)}\\
MD [eV]   & $-$0.083  &  $-$0.051  &  $-$0.228  &  $-$0.074  &     0.002  &  $-$0.097  &  $-$0.088  &  $-$0.151 \\
MAD [eV]  &    0.083  &     0.065  &     0.228  &     0.074  &     0.036  &     0.097  &     0.088  &     0.151 \\
MARD [\%] &    12.81  &      9.84  &     31.62  &     10.43  &      4.46  &     13.84  &     12.51  &     20.74 \\[2ex]
\multicolumn{9}{@{}l}{\it Hydrogen-bonded systems (13)}\\
MD [eV]   &    0.002  &  $-$0.019  &  $-$0.195  &  $-$0.082  &  $-$0.035  &  $-$0.177  &  $-$0.127  &  $-$0.160 \\
MAD [eV]  &    0.042  &  0.048     &  0.195     &  0.088     &  0.052     &  0.177     &  0.127     &  0.160    \\
MARD [\%] &    3.74   &  4.73      &  20.51     &  9.33      &  5.32      &  19.34     &  13.99     &  17.21    \\[2ex]
\multicolumn{9}{@{}l}{\it Average over all systems (23)}\\
MD [eV]   & $-$0.035  &  $-$0.033  &  $-$0.209  &  $-$0.079  &  $-$0.019  &  $-$0.142  &  $-$0.110  &  $-$0.156 \\
MAD [eV]  & 0.060     &  0.055     &  0.209     &  0.082     &  0.045     &  0.142     &  0.110     &  0.156    \\
MARD [\%] & 7.68      &  6.95      &  25.34     &  9.81      &  4.95      &  16.95     &  13.34     &  18.74    \\[5ex]
\multicolumn{9}{@{}l}{\boldmath{$\sqrt[3]{V}$}}\\
\multicolumn{9}{@{}l}{\it vdW-bonded systems (10)}\\
MD [\AA]  &    0.176  &     0.019  &  $-$0.105  &     0.035  &  $-$0.072  &  $-$0.169  &  $-$0.132  &  $-$0.139 \\
MAD [\AA] &    0.176  &     0.048  &  0.105     &     0.051  &  0.072     &  0.169     &  0.132     &  0.139    \\
MARD [\%] &    2.54   &     0.69   &  1.47      &     0.77   &  0.99      &  2.37      &  1.85      &  1.96     \\[2ex]
\multicolumn{9}{@{}l}{\it Hydrogen-bonded systems (13)}\\
MD [\AA]  &    0.176  &     0.045  &  $-$0.106  &  $-$0.010  &  $-$0.075  &  $-$0.175  &  $-$0.128  &  $-$0.119 \\
MAD [\AA] &    0.176  &     0.053  &  0.106     &  0.042     &  0.075     &  0.175     &  0.128     &  0.119    \\
MARD [\%] &    2.60   &     0.75   &  1.65      &  0.64      &  1.18      &  2.70      &  1.98      &  1.86     \\[2ex]
\multicolumn{9}{@{}l}{\it Average over all systems (23)}\\
MD [\AA]  &    0.176  &     0.034  &  $-$0.106  &     0.010  &  $-$0.074  &  $-$0.172  &  $-$0.130  &  $-$0.128 \\
MAD [\AA] &    0.176  &     0.051  &  0.106     &     0.046  &  0.074     &  0.172     &  0.130     &  0.128    \\
MARD [\%] &    2.57   &     0.72   &  1.57      &     0.70   &  1.10      &  2.56      &  1.92      &  1.90     \\\hline\hline
\end{tabular*}
\end{table*}

\begin{table*}[h!]\footnotesize
\caption{\label{table:Cu-benzene_energy}Adsorption distance [\AA]
defined as the carbon--metal distance, adsorption energy [eV], and their
relative deviation (RD) for benzene adsorbed on the (111) surface of Cu,
Ag, and Au. Reference data is available in
Refs.~\onlinecite{Berland_2009:rings_sliding,
Liu_2013:structure_energetics, Bilic_2006:adsorption_benzene,
Maass_2018:binding_energies, Peng_2016:versatile_vdW_SCAN-rVV10}. For
SCAN-VV we report the distance between the surface and the center of
mass of the benzene molecule.\cite{Peng_2016:versatile_vdW_SCAN-rVV10}}
\begin{tabular*}{\textwidth}{@{}l@{\extracolsep{\fill}}rrrrrrrrrr@{}}\hline\hline
System            &       expt.  &      DF1 &    DF2   &  DF1-optB88 &    DF1-cx  &    DF2-B86R &    DF3-opt1  &    DF3-opt2 &       VV  &   SCAN-VV \\\hline
\multicolumn{11}{@{}l}{\it Cu(111)/benzene}\\
distance [\AA]    &         2.83 &     3.60 &     3.49 &        3.34 &       2.95 &        3.10 &         2.87 &        3.01 &      3.13 &      2.93 \\
RD [\%]           &         ---  &    27.92 &    24.05 &       18.62 &       4.51 &       10.13 &         1.86 &        7.04 &     11.31 &      3.53 \\
energy [eV]       &0.68$\pm$0.04 &    0.464 &    0.430 &       0.595 &      0.708 &       0.530 &        0.646 &       0.617 &     0.697 &     0.740 \\
RD [\%]           &         ---  & $-$31.74 & $-$36.82 &    $-$12.52 &       4.06 &    $-$22.04 &      $-$4.95 &     $-$9.20 &      2.50 &      8.82 \\[2ex]
\multicolumn{11}{@{}l}{\it Ag(111)/benzene}\\
distance [\AA]    &         2.97 &     3.58 &     3.45 &        3.33 &       3.05 &        3.14 &         2.99 &        3.07 &      3.14 &      3.02 \\
RD [\%]           &         ---  &    20.44 &    16.05 &       12.24 &       2.64 &        5.61 &         0.54 &        3.28 &      5.85 &      1.68 \\
energy [eV]       &0.63$\pm$0.05 &    0.513 &    0.478 &       0.673 &      0.669 &       0.608 &        0.655 &       0.668 &     0.767 &     0.680 \\
RD [\%]           &         ---  & $-$18.63 & $-$24.10 &        6.82 &       6.20 &     $-$3.47 &         4.02 &        5.99 &     21.71 &      7.94 \\[2ex]
\multicolumn{11}{@{}l}{\it Au(111)/benzene}\\
distance [\AA]    &         3.05 &     3.50 &     3.36 &        3.28 &       3.06 &        3.12 &         3.01 &        3.06 &      3.11 &      3.07 \\
RD [\%]           &         ---  &    14.81 &    10.15 &        7.55 &       0.20 &        2.18 &         1.47 &        0.28 &      2.10 &      0.66 \\
energy [eV]       &0.71$\pm$0.03 &    0.601 &    0.575 &       0.771 &      0.767 &       0.653 &        0.724 &       0.712 &     0.839 &     0.730 \\
RD [\%]           &         ---  & $-$15.39 & $-$18.97 &        8.57 &       7.97 &     $-$8.00 &         1.96 &        0.21 &     18.11 &      2.82 \\\hline\hline
\end{tabular*}
\end{table*}

\end{appendix}

\clearpage\twocolumngrid
\section*{Associate Content}
\subsection*{Supporting Information}
The Supporting Information is available free of charge at
https://pubs.acs.org/doi/\dots

\begin{itemize}
\item Additional and more detailed computational results of the S22,
S22$\times$5, S66, S66$\times$8, G2--1, G2RC, and X23 datasets;
additional graphs of the exchange enhancement factor (XLSX)
\end{itemize}

\bibliographystyle{achemso}
\bibliography{references}

\clearpage\onecolumngrid
\section*{For Table of Contents Only}
\begin{center}
\includegraphics[width=3.25in]{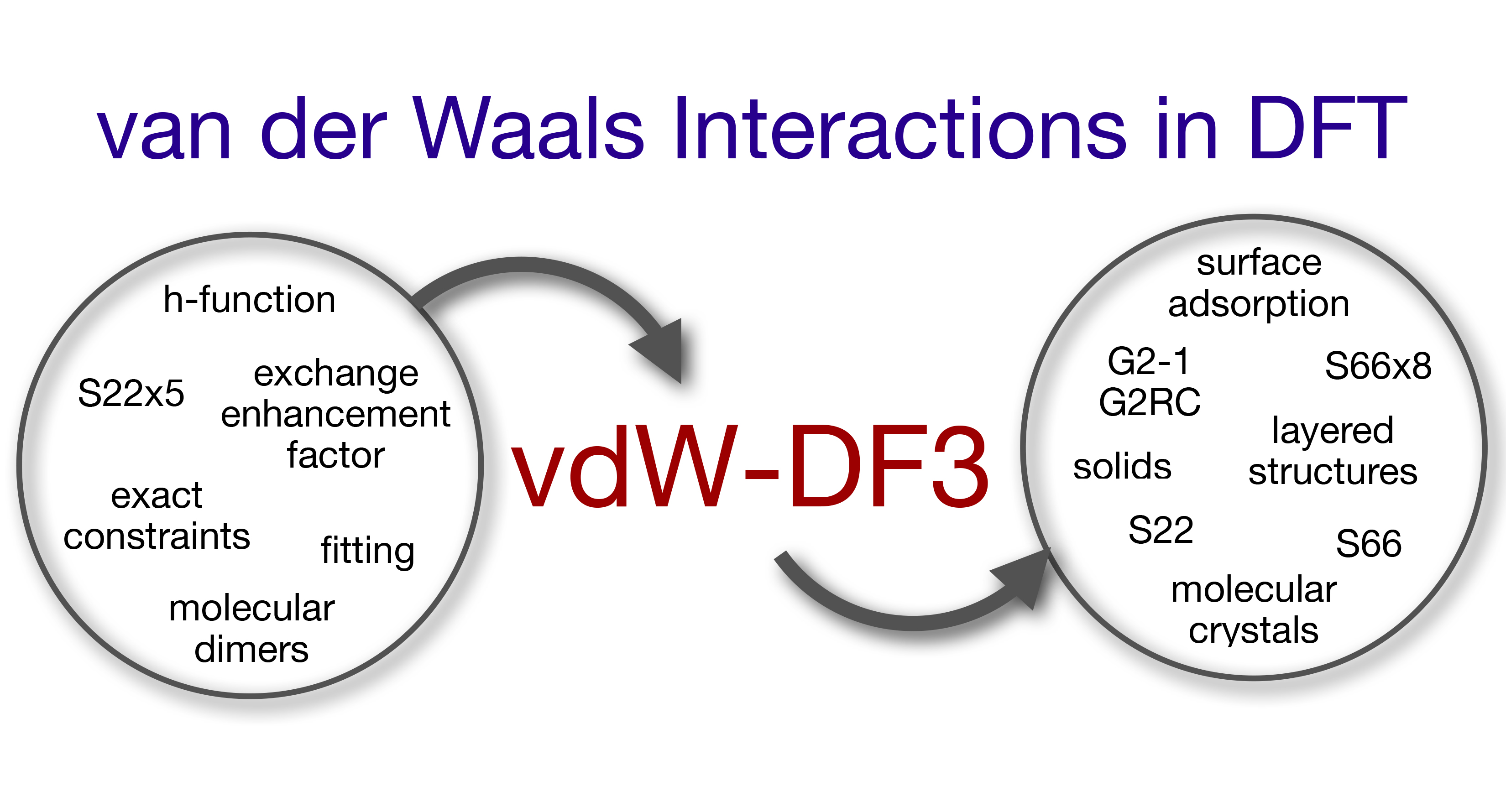}
\end{center}

\end{document}